\documentclass[review]{elsarticle}

\usepackage{lineno}
\usepackage[hidelinks]{hyperref}
\modulolinenumbers[5]
\journal{Applied Mathematics and Computation}

\newcommand{\domain}{\Omega}

\newcommand{\x}{\mathbf{x}}
\renewcommand{\v}{\mathbf{v}}
\newcommand{\q}{\mathbf{q}}

\newcommand{\jump}[1]{\left\llbracket #1 \right\rrbracket}

\newcommand{\Interface}{\Gamma}
\newcommand{\mi}{\dot m}

\newcommand{\sint}{S^\Gamma}
\newcommand{\dpi}{\Delta p_{\Gamma}}
\newcommand{\sprodi}{\eta^\Gamma}
\newcommand{\cc}{\sigma_c}

\newcommand{\lmm}{L_{mm}}
\newcommand{\lme}{L_{em}}
\newcommand{\lem}{L_{me}}
\newcommand{\lee}{L_{ee}}

\newcommand{\U}{\mathbf{U}}
\newcommand{\liq}{_\mathrm{liq}}
\newcommand{\vap}{_\mathrm{vap}}

\usepackage{amssymb}
\usepackage{amsmath}
\usepackage{amsthm}
\usepackage{stmaryrd} %
\usepackage{tikz}
\usepackage{pgfplots}
\usepackage{tabularx}
\usepackage{siunitx}

\begin{document}

\begin{frontmatter}

   \title{Riemann Solvers for Phase Transition in a Compressible Sharp-Interface Method}

\author{Steven Jöns}
\ead{joens@iag.uni-stuttgart.de}
\author{Claus-Dieter Munz}
\address{Institute of Aerodynamics and Gas Dynamics, University of Stuttgart, 70569 Stuttgart, Germany}

\begin{abstract}
   In this paper, we consider Riemann solvers with phase transition effects based on the Euler-Fourier equation system. One exact and two approximate solutions
   of the two-phase Riemann problem are obtained by modelling the phase transition process via the theory of classical irreversible thermodynamics. Closure is
   obtained by appropriate Onsager coefficients for evaporation and condensation. We use the proposed Riemann solvers in a sharp-interface level-set ghost fluid
   method to couple the individual phases with each other. The Riemann solvers are validated against molecular dynamics data of evaporating Lennard-Jones
   truncated and shifted fluid. We further study the effects of phase transition on a shock-drop interaction with the novel approximate Riemann solvers. 
\end{abstract}

\end{frontmatter}

\section{Introduction}
One of the most defining features of two-phase flows is phase transition. The interfacial transfer of mass, momentum and energy between two phases is able to strongly
influence flow fields and flow characteristics. Therefore, an accurate prediction of these effects is needed in many applications, e.g., weather forecast,
heat pipes or rocket engine combustion chambers. Depending on the mathematical description of two-phase fluid flow, different approaches to the
thermodynamic modelling of phase transition need to be taken. In general, one can identify two main approaches: diffuse- and sharp-interface methods.  In this
paper, we continue the work on our sharp-interface method \cite{Jons2021,Muller2020,Fechter2017} for compressible fluid flows. 

Therein, the interface is approximated as a discontinuity in the macroscopic fluid solution. The inclusion of phase transition effects is then confined to a
model of the interfacial transfer between the two phases, fulfilling appropriate jump conditions \cite{Ishii2011}. However, additional effort is
needed in order to track the movement of the interface and to identify, which part of the computational domain belongs to which fluid. We follow
\citet{Sussman1994} and use a level-set function to follow the interface. The fulfillment of the interfacial jump conditions is handled by a modification of the
Ghost Fluid method, which was introduced by Fedkiw et al. \cite{Fedkiw1999}. We apply a variant that was proposed by \citet{Merkle2007}, which relies on the use
of Riemann solvers to populate the ghost states that are needed to calculate numerical fluxes for each phase. This approach has been extended by
\citet{Fechter2018} for using  approximate Riemann solvers instead of the exact Riemann solution as used in \cite{Merkle2007}. In either case, an accurate
solution of the two-phase Riemann problem is the main building block in order to model the phase transition process correctly.

Riemann problems have been a pillar in the construction of numerical methods since the pioneering work of \citet{Godunov1959}. For the single phase case and
convex equations of state (EOS), the solution of the Riemann problem for the Euler equations is well-known and several approximate solutions exist to calculate
the numerical flux with less computational effort. For an overview the reader is referred to  the book of \citet{Toro_2009}. In the two-phase case, finding a
solution of the Riemann problem is a more intricate task. As \citet{Menikoff_1989} discussed, the non-convexity inside the spinodal region of the EOS leads to a
loss of hyperbolicity of the equations and thus an imaginary speed of sound. The authors attribute this behaviour to the EOS lacking physical information, which
leads to non-uniqueness of the solution.  Hence,  additional information about the physical process is needed. \citet{Menikoff_1989} assumed that the interface
is in thermodynamic equilibrium at any time and were able to discuss different wave patterns. The equilibrium assumption, however, is a strong premise as it is
in reality only applicable to infinitely slow phase transfer processes. In addition, it leads to a mixture of liquid and vapor as part of the solution. Related
works \cite{Saurel2008,LeMetayer2013,Furfaro2015,Kuila2015,Schwendeman2006} also treat the two-phase fluid as a mixture and are then able to formulate solutions
of
the Riemann problem. 

When keeping the fluids pure, the concept of a kinetic relation, proposed by \citet{Abeyaratne1991}, is a tool to circumvent the non-uniqueness of the Riemann
solution. Introducing a microscopic entropy measure across the interface allows to formulate a condition to identify an admissible solution. For the Euler
equations this was done by \citet{LeMetayer2005} when considering the reactive Riemann problem. This approach always chooses the solution with the maximal
possible entropy production. However, the Euler equations as mathematical model are not sufficient to describe phase transition as was shown by
\citet{Hantke2019}.  The authors argue and show that it is not possible to fulfill jump conditions in a thermodynamic consistent way in a case with phase
transition. They attributed this issue to the lack of a heat flux. Additional energy must be supplied to the interface in order to overcome the latent heat
of vaporization.  In the isothermal Euler equations, heat transfer is included implicitly with an infinite propagation rate. For this case, solutions of the
Riemann problem were discussed by \citet{Merkle2007}, \citet{Rohde2015} and
\citet{Hantke_2013}. 

The alternative is to supplement the Euler equations with a heat conduction term. This was considered by \citet{Fechter2017,Fechter2018} and
\citet{Thein2018}. However, these authors included a heat conduction term only at the interface and not in the bulk fluid. In \cite{Fechter2017,Fechter2018},
this lead to the appearance of an energetic source term  that was set to the latent heat of vaporization, defined at a reference temperature. This modification then allows a solution of the Riemann problem. However, this solution is not energy conserving as the employed source term adds energy to the fluid when phase
transition takes place. This error was recognized in \citet{Hitz2020}, where the heat conduction was not only considered at the interface but also in the bulk.
In addition, no source term was added at the interface, but interfacial heat fluxes from each side of the interface were considered, similar to the ideas of
\citet{Thein2018}.  With this approach, the fluid is in itself able to provide the energy, needed to balance the latent heat. 

The exact solution of the multiphase Riemann problem, proposed by \citet{Hitz2020}, was based on two subgrid models for the interfacial mass and heat fluxes. The mass flux model was derived as a
simple application of classical irreversible thermodynamic theory, which was also used to define the kinetic relation. It featured an empirical constant, which
was fitted against numerical reference data. The heat flux model was constructed from an exact solution of a heat equation, which was
assumed to hold across the interface. The exact solution of the Riemann problem was validated against molecular dynamics data showing a very good agreement for
density and velocity profiles but not so much in the temperature profile. 

In this paper, we follow up on the progress, made in \citet{Hitz2020}, and reconsider the thermodynamic modelling of phase transition in the two-phase
Riemann problem. Modelling the process as a coupled transport gives a thermodynamic prediction for both heat and mass flux. In addition, closure models
from, e.g. kinetic theory are available, relieving of the need of a coefficient fit. A similar approach was followed by \citet{Mueller2021} for the
Godunov-Peshkov-Romenski formulation of the continuum equations. In this formulation of the continuum description, heat transfer is modeled in a hyperbolic fashion, making the approach from \cite{Mueller2021} not directly 
transferable to the Euler-Fourier system that is considered in this paper. We incorporate our thermodynamic model into the exact solver proposed by \citet{Hitz2020}. In
addition, we formulate two approximate Riemann solvers, based on the work of \cite{Fechter2018,Mueller2021}. 

The paper is structured as follows: We begin in
section \ref{sec:fun} to describe the equation system as well as our thermodynamic modelling of phase transition. In section \ref{sec:num}, we shortly discuss
our sharp-interface method and continue in developing the modified exact Riemann solver as well as two approximate Riemann solvers. These are applied in some
numerical experiments in section \ref{sec:res}. We conclude in section \ref{sec:conc} with a short summary and an outlook.

\section{Fundamentals}\label{sec:fun}
Our description of two-phase flows is based on a sharp-interface approach. In this, we restrict ourselves to the consideration of an arbitrary fluid in a liquid
and a vapor phase, i.e., a single-component fluid. At each point $\x=(x,y,z)^T$ of the computational domain $\domain$, only one of the two phases exists. The
evolution of the two phases are each described by the set of conservation equations, which are discussed in section \ref{sec:EQS}. Surfaces in the domain, at
which the two phases are in contact with each other, are called interface, denoted by $\Interface$. This hypersurface is infinitesimally thin and carries no
mass nor energy. At the interface, the phases may exchange mass, momentum and energy. This exchange is based on the application of the conservation laws to
the interface, as well as the thermodynamic modelling, which is described in section \ref{sec:Evap}.

\subsection{Equations of Fluid Motion}\label{sec:EQS}
Each of the two phases shall be described as inviscid compressible fluids. According to the results \citet{Hantke2019}, we additionally need heat conduction,
since we want to consider evaporation under non-isothermal conditions. The full set of conservation equations are:
\begin{subequations} 
   \begin{gather}
      \rho_t + \nabla \cdot (\rho\v) =0,  \\
      (\rho \v)_t + \nabla \cdot (\rho \v \circ \v) + \nabla p  =0,\\
      (\rho e)_t + \nabla \cdot(\v( \rho e  + p ))   + \nabla \cdot \q=0.
   \end{gather}
\label{eq:EF}
\end{subequations}
Therein, $\rho$ describes the density, $\v=(u,v,w)^{\mathrm{T}}$ the velocity vector, $p$ the pressure, $\q$ the heat flux vector and $e$ the total energy per
unit mass. The latter is defined as the sum of the specific internal energy $\epsilon$ and kinetic energy:
\begin{equation}
   e=\epsilon + \frac 1 2 \v \cdot \v.
   \label{eq:energy}
\end{equation}
In order to close the system \eqref{eq:EF}, one needs to prescribe an EOS, linking the thermodynamic state variables density,
internal energy, and pressure with each other:
\begin{equation}
   p=p(\rho, \epsilon).
   \label{eq:EOS}
\end{equation}
Within our numerical framework, we are able to utilize real gas EOS like the Peng-Robinson EOS or state-of-the-art multi-parameter EOS from the well-known fluid library
\textit{CoolProp} in a highly efficient tabulation framework \cite{Foll2019}. Finally, the closure for heat conduction is obtained via Fourier's Law
\begin{equation}
   \q= -\lambda \nabla T,
   \label{eq:Fourier}
\end{equation}
with $\lambda $ denoting the thermal conductivity and $T$ the temperature. The resulting mathematical model is named Euler-Fourier system in the following. 

\subsection{Complete Evaporation at an Interface}  \label{sec:Evap}
The dynamics of an evaporating two-phase system is dominated by the interfacial exchange of mass, momentum and energy. From the macroscopic level of the fluid
flow, this exchange is described by interfacial jump conditions. In a reference system normal to the interface they read as
\begin{subequations}
   \begin{align}
      \jump{\mi}&=0, \label{eq:jumpmass} \\
      \mi\jump{u} + \jump{p} &= \dpi, \label{eq:jumpmomentum} \\
      \mi \jump{e} + \jump{u p} +\jump{q}&=\sint \dpi \label{eq:jumpenergy},
   \end{align}
\label{eq:jumps}
\end{subequations}
where $\mi=\rho(u-\sint)$ is the evaporation mass flux per unit area, $\dpi$ the increase of pressure due to surface tension based on the Young-Laplace law, and
$\sint$ the velocity of the interface.  The jump brackets are defined as $\jump{*}=*\vap  -*\liq$, with the subscript "$\mathrm{vap}$" denoting the vapor side and
"$\mathrm{liq}$" denoting the liquid side.

From the thermodynamic perspective, the evaporating interface can be described by the theory of non-equilibrium thermodynamics
\cite{Gyarmati1970,deGroot1984,Lebon2008,Kjelstrup2008}. In the following, we will only consider complete phase transition from one pure phase to another
pure phase. We will generally speak of evaporation. However, all considerations also apply to condensation. 

Focus of the thermodynamic modelling
lies on the fulfillment of the first and second laws of thermodynamics. The first law is part of the interfacial jump conditions, namely the energy jump condition
\eqref{eq:jumpenergy}. For the fulfillment of the second law, we formulate an entropy balance in addition to \eqref{eq:jumps}:
\begin{equation}
   \mi \jump{s} + \jump{\frac q T} = \sprodi.
   \label{eq:jumpentropy}
\end{equation}
Thereby, $s$ denotes the entropy per unit mass and $\sprodi$ the entropy production per unit area. In order to fulfill
the second law, 
\begin{equation}
   \sprodi\ge0
   \label{eq:secondlaw}
\end{equation}
must hold at any time.

Following \cite{Cipolla1974,Johannessen2006}, we write the entropy production of an evaporating interface as
\begin{equation}
   \sprodi = \mi \jump{-\frac g T + \frac{h\vap} { T}} + q\vap \jump{\frac 1 T}
   \label{eq:sprod}
\end{equation}
with $g$ denoting the Gibbs energy per unit mass and $h$ denoting the enthalpy per unit mass. Notably, Eq. \eqref{eq:sprod} is in general only
valid for stationary phase boundaries and under the assumption that kinetic energy terms do not contribute substantially to $\sprodi$.  For our following
considerations, the stationarity is easy to achieve by a simple transformation of the reference frame $\tilde u = u - \sint$. 

Starting from Eq. \eqref{eq:sprod}, phenomenological laws may be derived via Onsager theory for the two fluxes $\mi$ and $q\vap$. Therefore, one assumes a linear
relation between the fluxes and the two thermodynamic forces \cite{Lebon2008}, which are the terms inside the jump brackets:
\begin{align}
   \mi   = \lmm \jump{-\frac g T + \frac{h\vap} { T}} + \lme \jump{\frac 1 T}\label{eq:phenlawm}\\
   q\vap = \lem \jump{-\frac g T + \frac{h\vap} { T}} + \lee \jump{\frac 1 T}.
   \label{eq:phenlawq}
\end{align}
The coefficients $\lmm, \lme, \lem, \lee$ are so-called Onsager coefficients. In general, they are a function of the thermodynamic state, left and right of
the interface. 

The phenomenological laws stated above describe a coupled transfer phenomena, as each flux depends on both thermodynamic forces. For this process, we assume the Onsager
reciprocal relation to hold, i.e., $\lme= \lem$. Hence, to describe the non-equilibrium evaporation process, one needs to find closure models for three Onsager
coefficients from micro- or mesoscopic descriptions of the interface, e.g., \cite{Johannessen2006,Stierle2020}.

In the following, we will limit ourselves to the use of a closure model derived by \citet{Cipolla1974}, which is based on kinetic theory. In their work, the
authors derived expressions for microscopic temperature and density jumps and related these to the phenomenological laws described above. Their reported Onsager
coefficients are as follows:
\begin{gather}
   L_{mm}= \frac{-\nu_{2}}{\nu_{1}\nu_{2}-\nu_{3}^2} {\rho\vap \sqrt{\frac{2 T\liq}{R}}} \\
   L_{me}=L_{em}=\frac{-\nu_{3}}{\nu_{1}\nu_{2}-\nu_{3}^2}{\rho\vap T\liq \sqrt{{2 T\liq}}} \\
   L_{ee}=-\frac{\nu_{1}}{\nu_{1}\nu_{2}-\nu_{3}^2}{p_s(T\liq) T\liq \sqrt{{2 T\liq}}} \\
\end{gather}
with the definitions
\begin{gather}
   \nu_1=\frac 9 8 \sqrt \pi \left( \frac 1 2 + \frac{16}{9\pi}\right)-\sqrt \pi \frac {1-\cc}{\cc} \\
   \nu_2=\frac 4 8 \sqrt \pi \left( \frac 1 2 + \frac{52}{25\pi}\right) \\
   \nu_3=\frac 2 8 \sqrt \pi \left( \frac 1 2 + \frac{8}{5\pi}\right)
\end{gather}
where $p_s(T\liq$) is the saturation pressure evaluated at the temperature $T\liq$ and $\cc$ the so-called condensation coefficient. It accounts for the fact
that not every liquid molecule that reaches the interface will evaporate. To close the entire evaporation model we use the formulation of \citet{Nagayama2015}
to describe the evaporation coefficient.

\section{Numerical Method}\label{sec:num}
Sharp-interface simulations of two-phase flow are based on three essential building blocks: A basic numerical scheme for the bulk phases,  a tracking of the
interface and a coupling algorithm for the individual bulk phases at the interface. Our approach is a continuation of the work of Fechter
\cite{Fechter2017,Fechter2015b}, for which a detailed description can be found in \cite{Jons2021,Mueller2021,Zeifang2020}. In the following, we give a short
overview of the numerical method to keep the paper self-consistent. First, we discuss the discretization of the fluid flow equations. Afterwards, the intricate
parts of two-phase flow simulations are presented: the tracking of the interface and the coupling of the individual phases by the solution of the Riemann
problem. Finally, we present in detail the main result of this paper, namely three different Riemann solvers taking into account phase changes.

\subsection{Discretization of Fluid Flow Equations}\label{sec:DGSEM}
We solve the fluid flow equations \eqref{eq:EF} with the discontinuous Galerkin spectral element (DGSEM) flow solver
\textit{FLEXI}\footnote{https://www.flexi-project.org/} \cite{Krais2020}. Therein, the computational domain is discretized into $N_{Elems}$ non-overlapping
elements. In each element, the fluid flow equations are rewritten into a weak formulation. We approximate the solution in the basis of Lagrange polynomials of  degree $N$,
in which the interpolation points are defined to be the Gauss points. The individual elements of the grid are coupled with each other by employing well-known Riemann solvers for single phase flow 
at the element boundaries. Gradients for the parabolic fluxes are obtained with the BR1 lifting operator applied to the primitive variables
\cite{Bassi2011,Bassi2002}. 

While the DGSEM leads to high order accurate solutions for smooth flows, discontinuities like shock waves or interfaces ultimately lead to well-known Gibbs
oscillations. Therefore, non-smooth phenomena like shock waves and contact discontinuities are captured via finite volume sub-cells \cite{Sonntag2014,Sonntag2017}. To
alleviate the loss of the order of accuracy, the underlying finite-volume scheme on the sub-grid is constructed as a second order TVD scheme with a minmod limiter. As in the DG method,
numerical fluxes are calculated via Riemann solvers. Solution gradients for the heat conduction at the sub-cell interfaces are obtained with the Gauss-Green method. The transformation
between polynomial solution and the finite volume sub-cells is treated by enforcing integral conservation, while the decision to switch between each formulation
is done based on the modal smoothness indicator of \citet{Persson2006}.  Finally, temporal evolution of the weak formulation for both DG and FV formulations is
done with a fourth order low-storage Runge-Kutta scheme \cite{Carpenter1994}. 

\subsection{Interface Tracking \& Ghost Fluid Method}\label{sec:si}
We track the position of the phase interface by a level-set function $\phi(\x)$ that is advected following \cite{Sussman1994} by 
\begin{equation}
   \frac{\partial \phi}{\partial t} + \v^{LS} \nabla \phi =0,
   \label{eq:ls}
\end{equation}
where $\v^{LS}=(u^{LS},v^{LS},w^{LS})^{\mathrm{T}}$ is the level-set advection velocity field. Its value is in principle known only at the interface from the solution of the
two-phase Riemann Problem (see section \ref{sec:Riemann}). We extrapolate this field into the volume via the method of \citet{Peng1999} with a 5th order WENO scheme
\cite{Jiang2000}.
We solve Eq.\eqref{eq:ls} with a path-conservative DGSEM \cite{castro2006high,Dumbser2016a} as discussed in \cite{Jons2021}.
Time integration is done with the same method as in the bulk fluids. The level-set function is initially a signed-distance function. However, due to the nature
of the transport governed by a compressible fluid, the signed-distance property is not upheld for all times. Therefore, we utilize the reinitialization procedure
proposed by \citet{Peng1999} and  solve the reinitialization equation with the same WENO scheme as in the extrapolation procedure.

\begin{figure}
   \centering
   \begin{tikzpicture}
      \draw[black,thick,->] (-0.2,0)--(8.2,0) node [right]{$x$};
      \draw[black,dashed,thick]  (0,-1)--(0,2);
      \draw[black,dashed,thick]  (2,-1)--(2,2);
      \draw[red,dashed,thick]    (4,-1)--(4,2) node[above] {$\Gamma$};
      \draw[black,dashed,thick]  (6,-1)--(6,2);
      \draw[black,dashed,thick]  (8,-1)--(8,2);
      \draw[black]  (1,-0.05)--(1,0.05);
      \draw[black]  (3,-0.05)--(3,0.05);
      \draw[black]  (5,-0.05)--(5,0.05);
      \draw[black]  (7,-0.05)--(7,0.05);
      \node[below] at  (1,0){$x_{i-1}$};
      \node[below] at  (3,0){$x_{i  }$};
      \node[below] at  (5,0){$x_{i+1}$};
      \node[below] at  (7,0){$x_{i+2}$};
       
      \draw[blue,thick,->]  (-0.2,1)--(0.2,1);
      \draw[blue,thick,->]  ( 1.8,1)--(2.2,1);
      \draw[red, double,->]  ( 3.6,1.1)--(4.0,1.1);
      \draw[red, double,->]  ( 4.0,0.9)--(4.4,0.9);
      \draw[blue,thick,->]  ( 5.8,1)--(6.2,1);
      \draw[blue,thick,->]  ( 7.8,1)--(8.2,1);
      \draw[red, double,->]  ( 0.0,-1.6)--(0.4,-1.6) node [right,black] {Two-Phase Riemann solver};
      \draw[blue,thick,->]( 0.0,-2.0)--(0.4,-2.0) node [right,black] {Classical Riemann solver};
   \end{tikzpicture}
   \caption{Flux calculation of the Riemann solver based Ghost Fluid method. In the pure bulk phases, classical Riemann solvers are used. At the
  interface, lying between $x_i$ and $x_{i+1}$, numerical fluxes are calculated for each phase based on the solution of a two-phase Riemann problem.}
   \label{fig:Ghost}
\end{figure}
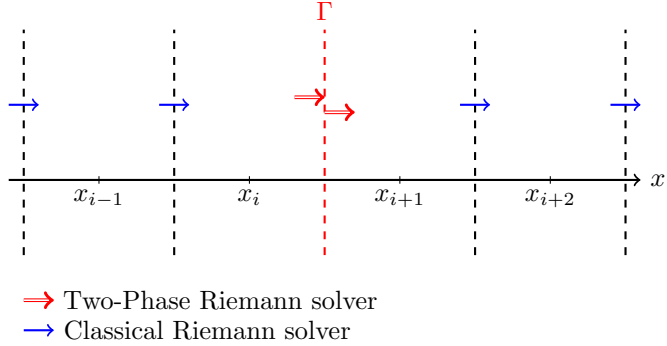
From the sign of the level-set field, the phase of the approximate solution at each point of the computational domain can be identified. With this information,
time evolution of the bulk phases is straightforward, given the computational method described in section \ref{sec:DGSEM}. The  information about a change of
sign of the level-set function is also used to switch from a DGSEM element to a FV sub-grid to improve the localization of the interface and to increase the
overall stability. For the coupling at the interface we follow the work of \citet{Merkle2007}, who extended the Ghost Fluid method by \citet{Fedkiw1999}. The
idea of the Ghost Fluid method is to define two states in grid cells adjacent to the interface: a real state and a ghost state. The interface tracking then
defines, which of these states has to be taken to determine the numerical flux at the interface for the bulk solver. \citet{Merkle2007} used information from
the local solution, as given by the solution of the interface Riemann problem, to define the ghost state. The Riemann problem at the interface also provided 
the local velocity of the interface that is needed for the level-set advection equation. Hence, at the interface the classical Riemann solvers are
exchanged by a special two-phase
Riemann solver in order to correctly couple the individual phases. The numerical flux for the finite volume sub-cell is calculated for each phase individually,
as will be discussed in section \ref{sec:Riemann}. The procedure is visualised in Fig. \ref{fig:Ghost}.

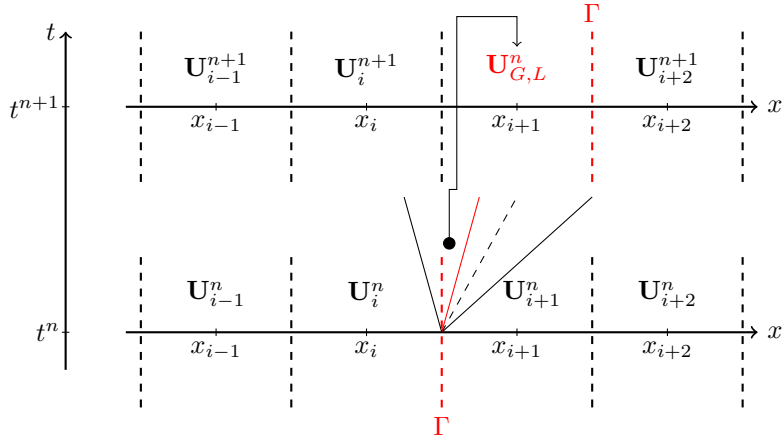
\begin{figure}
   \usetikzlibrary{arrows}
   \centering
   \begin{tikzpicture}
      \draw[black,thick,->] (-0.2,0)--(8.2,0) node [right]{$x$};
      \draw[black,thick,->] (-0.2,3)--(8.2,3) node [right]{$x$};
      \draw[black,thick,->] (-1.,-0.5)--(-1.,4) node [left]{$t$};
      \draw[black,dashed,thick]  (0,-1)--(0,1);
      \draw[black,dashed,thick]  (2,-1)--(2,1);
      \draw[red,dashed,thick]    (4, 1)--(4,-1) node[below] {$\Gamma$};
      \draw[black,dashed,thick]  (6,-1)--(6,1);
      \draw[black,dashed,thick]  (8,-1)--(8,1);
      \draw[black,dashed,thick]  (0, 2)--(0,4);
      \draw[black,dashed,thick]  (2, 2)--(2,4);
      \draw[black,dashed,thick]  (4, 2)--(4,4);
      \draw[red,dashed,thick]    (6, 2)--(6,4) node[above] {$\Gamma$};
      \draw[black,dashed,thick]  (8, 2)--(8,4);
      
      \draw[black]  (-1.05,0)--(-0.95,0) node[left] {$t^n$};
      \draw[black]  (-1.05,3)--(-0.95,3) node[left] {$t^{n+1}$};

      \draw[black]  (1,-0.05)--(1,0.05);
      \draw[black]  (3,-0.05)--(3,0.05);
      \draw[black]  (5,-0.05)--(5,0.05);
      \draw[black]  (7,-0.05)--(7,0.05);
     
      \draw[black]  (1,3-0.05)--(1,3.05);
      \draw[black]  (3,3-0.05)--(3,3.05);
      \draw[black]  (5,3-0.05)--(5,3.05);
      \draw[black]  (7,3-0.05)--(7,3.05);

      \node[below] at  (1,0){$x_{i-1}$};
      \node[below] at  (3,0){$x_{i  }$};
      \node[below] at  (5,0){$x_{i+1}$};
      \node[below] at  (7,0){$x_{i+2}$};
 
      \node[below] at  (1,3){$x_{i-1}$};
      \node[below] at  (3,3){$x_{i  }$};
      \node[below] at  (5,3){$x_{i+1}$};
      \node[below] at  (7,3){$x_{i+2}$};

      \node[] at  (1,0.5){$\U_{i-1}^n$};
      \node[] at  (3,0.5){$\U_{i  }^n$};
      \node[] at  (5.2,0.5){$\U_{i+1}^n$};
      \node[] at  (7,0.5){$\U_{i+2}^n$};
      
      \node[] at  (1  ,3.5){$\U_{i-1}^{n+1}$};
      \node[] at  (3  ,3.5){$\U_{i  }^{n+1}$};
      \node[red] at  (5.0,3.5){$\U_{G,L}^{n}$};
      \node[] at  (7  ,3.5){$\U_{i+2}^{n+1}$};

      \draw[]         (4,0)--(6  ,1.8);
      \draw[dashed]   (4,0)--(5  ,1.8);
      \draw[red]      (4,0)--(4.5,1.8);
      \draw[]         (4,0)--(3.5,1.8);

      \draw[*->] (4.1,1.1) --(4.1,1.9)--(4.2,1.9)--(4.2,4.2)--(5,4.2)--(5,3.8);
   \end{tikzpicture}
   \caption{Procedure of setting the ghost state in the Ghost-Fluid method. If the level-set advection moves the interface from one sub-cell side to another,
   the phase of the respective degree of freedom changes ($i+1$). The conservative state vector in that cell is then populated with the interfacial state from the two-phase
Riemann problem of the respective phase ($\U^n_{G,L}$).}
   \label{fig:Ghoststate}
\end{figure}

Special treatment is needed if the interface moves across the mesh. This case is depicted in Fig \ref{fig:Ghoststate}. Whenever the advection shifts the
interface from one sub-cell side to another, the cell changes its phase. Therefore, we redefine the state in the corresponding sub-cell by the
ghost state of the same phase. The ghost states are the inner values of the Riemann solution  adjacent to the interface and are described in the next subsection.
This procedure is non-conservative but allows for highly complex two-phase flow simulations while maintaining a simple fixed, computational grid
\cite{Jons2021,Fechter2017,Mueller2021,Zeifang2020}. We can circumvent this step in a one-dimensional setting by enforcing that the interface is always aligned
with the same sub-cell side. This can be done following Hitz et al.  \cite{Hitz2020,Hitz2020a}, where the DGSEM is supplemented with an
Arbitrary-Lagrangian-Eulerian (ALE) formulation \cite{Minoli2011}. With this, we can move the mesh with the velocity field $S^M=u^{LS}$ and the need for setting
the ghost state vanishes.  

\subsection{Exact and Approximate Two-Phase Riemann Solvers}\label{sec:Riemann}
\begin{figure}
   \centering
   \begin{tikzpicture}
      \draw[->,thick] (0,0)--(0,4) node[left] {$t$};
      \draw[->,thick] (-4,0)--(4,0) node[below] {$x$};
      
      \draw[thick]  (0,0) -- (4,3.5);
      \draw[thick,dashed]  (0,0) -- (1,3.5);
      \draw[thick,dotted]  (0,0) -- (-0.5,3.5);
      \draw[thick]  (0,0) -- (-2.3,3.5);
      \draw[thick]  (0,0) -- (-2.6,3.5);
      \draw[thick]  (0,0) -- (-2.9,3.5);

      \node at (-2,1) {$\U\liq$};
      \node at (-1,2.2) {$\U\liq^*$};
      \node at ( 1.6,2.2) {$\U\vap^*$};
      \node at ( 0.4,3.0) {$\U\vap^\sharp$};
      \node at ( 2,1) {$\U\vap$};
   \end{tikzpicture}
   \caption{Wave pattern of the two-phase Riemann Problem. The shock wave is depicted as a solid line, the contact discontinuity as dashed line, the interface
   as a dotted line and the rarefaction wave as a fan of solid lines.}
   \label{fig:tpR}
\end{figure}
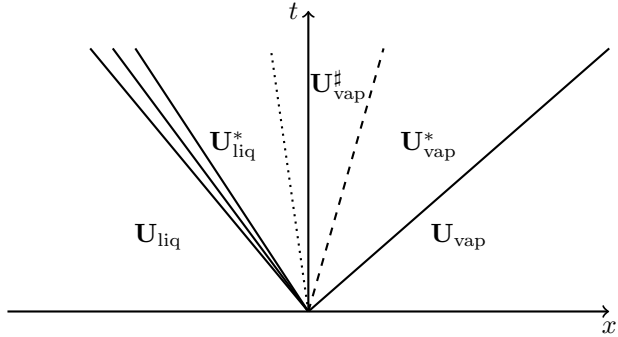

The remaining building block of our numerical method is the two-phase Riemann Problem and its solution. We define the two-phase Riemann Problem as the initial value
problem with initially a liquid state on the left and a vapor state on the right. Similar to the classical Riemann Problem, waves emerge from the initial
discontinuity for $t>0$ as depicted in Fig. \ref{fig:tpR}. These waves include classical waves like shock, rarefaction and contact discontinuity, as well as a
non-classical wave: the phase interface. Across this wave, mass, momentum and energy is exchanged between the two phases depending on the characteristics of the
phase transition at play.

The equation system \eqref{eq:EF} makes the consideration of the Riemann problem difficult due to the inclusion of the heat flux.  Its dissipative nature  will
essentially lead to the wave speeds reducing over time. Consequently, the states between the waves are not constant in space or time, leading to a loss of
self-similarity of the solution. All simple solution strategies for obtaining Riemann problem solutions fail in this case. In order to keep the Riemann problem
solvable, we simplify it without removing any of its key aspects. We assume that the solution is self-similar as shown in Fig. \ref{fig:tpR}. In addition, heat
fluxes shall be negligible across all waves, except for the interface. The interfacial heat fluxes are constant in time and need to be accounted for in the
calculation of the numerical flux. In other words, the heat flux is considered in the jump conditions at the evaporation front only. We note that this
simplification is only introduced in the multiphase Riemann solver.

An additional difficulty is brought to the solution by the phase interface. This undercompressive shock wave is not associated with any eigenvalue of the
equation system.  The appearance of such non-classical wave structures in Riemann Problems  was discussed in length by \citet{Menikoff_1989}. It is an underlying
issue of the equation of state, which is non-convex inside the spinodal region. Hence, for this unstable region of the EOS, the governing equations loose
hyperbolicity and the speed of sound becomes imaginary. The authors argued that this is a sign of  missing physical information, ultimately leading to
non-uniqueness of the solution. This in turn can not be treated by simple physical admissibility criterions like the Lax or Liu entropy criterions and more
advanced considerations are needed. In the following, we discuss three different approaches to solve the two-phase Riemann problem in order to calculate
numerical fluxes: one exact solver and two approximate Riemann solvers.

\subsubsection{Exact Riemann Solver}
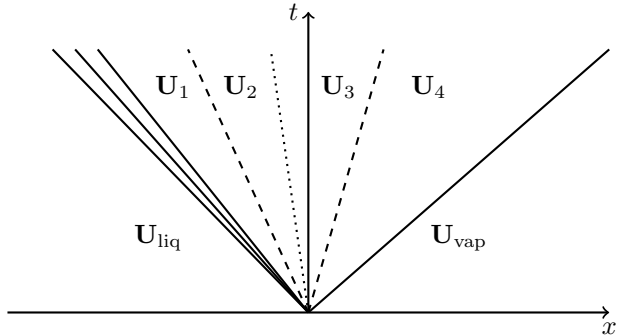
\begin{figure}
   \centering
   \begin{tikzpicture}
      \draw[->,thick] (0,0)--(0,4) node[left] {$t$};
      \draw[->,thick] (-4,0)--(4,0) node[below] {$x$};
      
      \draw[thick]  (0,0) -- (4,3.5);
      \draw[thick,dashed]  (0,0) -- (1,3.5);
      \draw[thick,dotted]  (0,0) -- (-0.5,3.5);
      \draw[thick,dashed]  (0,0) -- (-1.6,3.5);
      \draw[thick]  (0,0) -- (-2.8,3.5);
      \draw[thick]  (0,0) -- (-3.1,3.5);
      \draw[thick]  (0,0) -- (-3.4,3.5);

      \node at (-2,1) {$\U\liq$};
      \node at (-1.8,3.0) {$\U_1$};
      \node at ( -0.9,3.0) {$\U_2$};
      \node at ( 0.4,3.0) {$\U_3$};
      \node at ( 1.6,3.0) {$\U_4$};
      \node at ( 2,1) {$\U\vap$};
   \end{tikzpicture}
   \caption{Numerical wave pattern of the exact two-phase Riemann solver. The shock wave is depicted as a solid line, the contact discontinuities as dashed
      line, the interface as a dotted line and the rarefaction wave as the fan of solid lines.}
   \label{fig:exact}
\end{figure}

The exact Riemann solver is a modification of the solver of \citet{Hitz2020}. As depicted in Fig.\ref{fig:exact}, it assumes a similar wave structure of the
solution as shown in Fig. \ref{fig:tpR}, only supplemented by an additional contact discontinuity. This additional contact wave is introduced for numerical
reasons to allow evaporation as well as condensation during the iteration. It turned out that this introduction strongly increased the robustness of the
iteration procedure. Within the final solution, the additional contact wave drops out. For details of this concept see \cite{Fechter2017,Hitz2020,Zeiler2015}. A
solution for this type of wave pattern can be achieved by postulating Rankine-Hugoniot jump conditions across shock waves and contact discontinuities, isentropic
relations across rarefaction waves as well the two phase jump conditions \eqref{eq:jumps} across the phase interface. However, this is not sufficient to
eliminate the non-uniqueness problems discussed above.

To treat this issue, we follow the works of \cite{Merkle2007,Rohde2015,Hitz2020} and employ the concept of a kinetic relation, initially proposed by
\citet{Abeyaratne1991}. The idea is to define a function of the states left and right of the interface, that controls the amount of entropy production across
the wave. The authors in \cite{Merkle2007} write the kinetic relation as
\begin{equation}
   \mathcal{ \tilde{ K}} = \sum_{i=1}^{n_f}( f_i \tilde J_i) - G (\dot J_1,\cdots,\dot J_{n_f}),
   \label{eq:kineticrelationrohde}
\end{equation}
with $f_i$ denoting the independent thermodynamic forces,  $\tilde J_i$ the corresponding fluxes calculated from the macroscopic situation during the current
iteration, $\dot J_i$ the same fluxes calculated from a microscale model, and $n_f$ the total number of independent fluxes. The left term in
Eq. \eqref{eq:kineticrelationrohde} describes the amount of entropy production based on the iterated macroscopic quantities whereas $G$ is a microscopic
prediction. 

A possible formulation of the kinetic relation can be defined by combining Eq. \eqref{eq:kineticrelationrohde} with the evaporation model discussed in section
\ref{sec:Evap}. This approach would explicitly compare the entropy production of the current iterated solution with the entropy production predicted by the
evaporation model. However, the corresponding equation is highly nonlinear and was found to lack robustness when used in the proposed Riemann solver. 
Therefore,  we divert from the works of \cite{Merkle2007,Rohde2015,Hitz2020} in our choice of the kinetic relation and write it similar to \citet{Mueller2021} as
\begin{equation}
   \mathcal K = \tilde m^k - \dot m^k, 
   \label{eq:kineticrelation}
\end{equation}
where the superscript $k$ describes the current iteration step, $\tilde m$ the mass flux defined by the current iterated state variables and $\dot m$ the mass
flux defined from Eq. \eqref{eq:phenlawm}. Eq. \eqref{eq:kineticrelation} only enforces that the iterated mass flux is identical to the one predicted by the
evaporation model. In order to ensure the correct amount of entropy production at the interface, we further prescribe the vapor heat flux directly from the
phenomenological law \eqref{eq:phenlawq}. Thus, in a converged state, for which $\mathcal K\approx 0$ holds, entropy production at the interface fulfills Eq.
\eqref{eq:sprod} and therefore Eq. \eqref{eq:kineticrelationrohde} as well.

The full equation system that needs to be solved in order to obtain a Riemann solution consists of eight equations. For the iterative procedure we formulate these in the following as residual equations. With nomination of the intermediate states as depicted in Fig.\ref{fig:exact}, the equations related to the classical waves read as
\begin{align}
   r_1&=
   \begin{cases}
      s_1-s\liq &\,\text{if} \quad p_1<p\liq \\
      \epsilon_1-\epsilon\liq + \frac 1 2 (p\liq+p_1)(\tau_1-\tau\liq) &\, \text{if} \quad p_1\geq p\liq
   \end{cases},\\
   r_2&=
   \begin{cases}
      s_4-s\vap &\,\text{if} \quad p_4<p\vap \\
      \epsilon\vap-\epsilon_4 + \frac 1 2 (p_4+p\vap)(\tau\vap-\tau_4) &\, \text{if} \quad p_4\geq p\vap
   \end{cases}, \\
   r_3&=p_2-p_1, & \\
   r_4&=p_4-p_3, & \\
   r_5&=
   \begin{cases}
      T_4-T_3 &\,            \text{if} \quad \tilde{m}<0 \\
      (T_2-T_1)(T_4-T_3) &\, \text{if} \quad \tilde{m}=0 \\
      T_2-T_1 &\,            \text{if} \quad \tilde{m}>0
   \end{cases},
\end{align}
where $\tau$ denotes the specific volume and the mass flux $\tilde m$ is defined by
\begin{equation}
   \tilde m =\frac{u_3-u_2}{1/\rho_3-1/\rho_2}.
\end{equation}
Furthermore, the residuals pertaining to the interface are
\begin{align}
   r_6&=\tilde m(u_3-u_2)+p_3-p_2 + \Delta p_{\sigma}, \\
   r_7&=\tilde m(\epsilon_3-\epsilon_2 + 0.5(u_3^2-u_2^2))+p_3u_3-p_2u_2 + q\vap-q\liq -\dpi \sint, \\
   r_8&=\mathcal K.
\end{align}
The difference between the procedure considered here and the work of Hitz et al. \cite{Hitz2020,Hitz2020a} lies in the residuals $r_7$ and $r_8$. The residual $r_7$
refers to the energy jump across the interface. In contrast to Hitz et al. \cite{Hitz2020,Hitz2020a} we use a different but equivalent formulation of the jump
condition. In addition, the two interfacial heat fluxes are not defined by the subgrid model proposed in \cite{Hitz2020,Hitz2020a}. They are rather a direct
consequence of the evaporation model. The vapor heat flux is therefore given by Eq. \eqref{eq:phenlawq}. The liquid heat flux, can be calculated by considering
the energy jump condition from the micro-scale:
\begin{equation}
   q\liq=\dot m(\epsilon_3-\epsilon_2 + 0.5(u_3^2-u_2^2))+p_3u_3-p_2u_2 + q\vap -\dpi \sint.
   \label{eq:ejmicro}
\end{equation}
Note that the difference between Eq. \eqref{eq:ejmicro} and the residual $r_7$ lies in the use of the mass fluxes $\dot m$ and $\tilde m$, respectively. Once
the kinetic relation is fulfilled, both equations are identical. 

The numerical solution of this problem is obtained by using the \textit{GNU Scientific Library} \cite{Galassi} and its FORTRAN implementation FGSL. We use the
multi-dimensional Newton algorithm included in the library. The iterative solver is converged once all eight residuals fall under a user-defined threshold:
\begin{equation}
   \textbf{G}_{TRP}=(r_1,r_2,r_3,r_4,r_5,r_6,r_7,r_8)^\mathrm{T}< \Delta_{TRP}.
   \label{eq:goalfunctionexxact}
\end{equation}
From the exact solution, we calculate numerical fluxes following Godunov \cite{Godunov1959}, by evaluating the flux left and right of the interface:
\begin{align}
   F^*\liq=F(\U_2)+ (0,0,q\liq)^\mathrm{T} - S^M \U_{2} ,\label{eq:nfexactliq}\\
   F^*\vap=F(\U_3)+ (0,0,q\vap)^\mathrm{T} - S^M \U_{3} ,\label{eq:nfexactvap}
\end{align}
with $F(\U)$ being the convective flux function. Note that the second and third term in Eqs. \eqref{eq:nfexactliq} and \eqref{eq:nfexactvap} refer to the heat
fluxes defined from the evaporation model, as well as the ALE contribution. The latter vanishes when the mesh is fixed.

\subsubsection{HLLP Approximate Riemann Solver}
The remedies of the exact Riemann solver are obvious. The cumbersome procedure, described before, is in need of an extensive amount of computational effort. For
multi-dimensional applications, this is beyond feasible. In addition, robustness is limited, as during the iteration process, one can not guarantee that a guess does not 
lie inside the unstable spinodal region of the EOS, ultimately leading to an abortion of the algorithm. 

\begin{figure}
   \centering
   \begin{tikzpicture}
      \draw[->,thick] (0,0)--(0,4) node[right] {$t$};
      \draw[->,thick] (-4,0)--(4,0) node[below] {$x$};
      
      \draw[thick]  (0,0) -- (4,3.5) node [above] {$S\vap$};
      \draw[thick,dotted]  (0,0) -- (-0.5,3.5)node [above] {$\sint$};
      \draw[thick]  (0,0) -- (-3.4,3.5)node [above] {$S\liq$};

      \node at (-2,1) {$\U\liq$};
      \node at (-1.8,3.0) {$\tilde\U\liq^*$};
      \node at ( 1.6,3.0) {$\tilde\U\vap^*$};
      \node at ( 2,1) {$\U\vap$};
   \end{tikzpicture}
   \caption{Numerical wave pattern of the approximate HLLP two-phase Riemann solver. The outer waves are depicted as a solid line and the interface as a dotted
   line. The inner states are defined by the extended state vector $\tilde U$.}
   \label{fig:HLLP}
\end{figure}
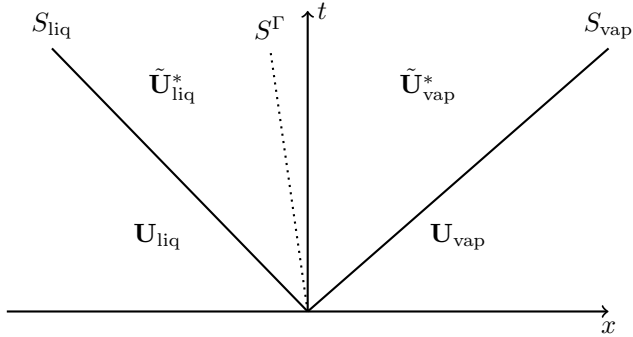

Therefore, we further propose a simpler and more robust approximate Riemann solver based on the idea of the HLLP solver of \citet{Fechter2018}. In their work, the authors
formulate an approximate Riemann solver for the two-phase Riemann problem building on the HLL \cite{Harten_1983} and HLLC \cite{Toro_1994,Hu2009} Riemann
solvers. The solution is assumed to oblige the simplified wave fan depicted in Fig. \ref{fig:HLLP}. Two outer, non-linear waves enclose the two spatially
constant states $\tilde \U\liq^*$ and $\tilde \U\vap^*$, that are separated by the phase interface. The two inner states are defined by an extended state vector
$\tilde \U^*=(\rho^*,\rho^*u^*,\rho^*v^*,\rho^*w^*,\rho^*e^*, p^*)^{\mathrm{T}}$. They are thermodynamically overdetermined, e.g.,  the quantities $\rho^*$,
$e^*$ and $p^*$  are generally not consistent with the EOS. This aspect is not a problem in the single phase \cite{Toro_1994} and non-evaporating two-phase
formulation \cite{Hu2009} of the HLLC solver. However, when considering evaporation, this aspect needs to be kept in mind.

\citet{Fechter2018} formulated their HLLP solver in an iterative fashion with the interface velocity $\sint$ being the iteration variable. Given an appropriate
approximation of the outer waves and a guess for $\sint$ the authors calculated the inner states by applying the jump conditions. As destination function the
authors used a kinetic relation which was evaluated with the conservative part of the inner states
${\U}^*=(\rho^*,\rho^*u^*,\rho^*v^*,\rho^*w^*,\rho^*e^*)^{\mathrm{T}}$. A solution was obtained once the kinetic relation was fulfilled. Notably, their
approach was only possible by removing the interfacial energy jump condition from the total set of jump
conditions, as the equation system for the inner states was overdetermined.

In this work, we follow the principal idea of the authors in \cite{Fechter2018} and formulate an HLLP solver based on the evaporation model discussed in section
\ref{sec:Evap}. A similar adaption was proposed by \citet{Mueller2021} for the Godunov-Peshkov-Romenski equation system. We start our approximate Riemann solver
by a two-phase adapted estimate of Davis \cite{Davis1988} for the outer wave speeds:
\begin{equation} 
   S\liq=u\liq-a\liq \quad \text{and} \quad S\vap=u\vap+a\vap,
 \label{eq:wavespeeds} 
 \end{equation}
where $a$ refers to the speed of sound. For a more compact form of the equations, we divert from \citet{Fechter2018} and follow \citet{Mueller2021} and choose
the mass flux across the interface $\tilde m $ as iteration variable.  We define the following boundary conditions from the values of the left and right initial state by applying the Rankine-Hugoniot jump conditions 
\begin{gather} 
    m\liq=\rho\liq(u\liq-S\liq)=\rho\liq^*(u\liq^*-S\liq),\label{eq:ml} \\
    m\vap=\rho\vap(u\vap-S\vap)=\rho\vap^*(u\vap^*-S\vap),\label{eq:mr}
\end{gather}
where $m\liq$ and $m\vap$ are the net mass fluxes across the outer waves, 
\begin{gather}
 I\liq=m\liq u\liq + p\liq=m\liq u\liq^* + p\liq^*, \label{eq:il} \\
 I\vap=m\vap u\vap + p\vap=m\vap u\vap^* + p\vap^*, \label{eq:ir} 
\end{gather}
$I\liq$ and $I\vap$ the net momentum fluxes across the outer waves,
\begin{gather}
   E\liq=m\liq e\liq + p\liq u\liq=m\liq e\liq^* + p\liq^* u\liq^*, \label{eq:el} \\
   E\vap=m\vap e\vap + p\vap u\vap=m\vap e\vap^* + p\vap^* u\vap^*, \label{eq:er} 
\end{gather}
and $E\liq$ and $E\vap$ the net energy fluxes across the outer waves. Inserting the definitions \eqref{eq:il} and \eqref{eq:ir} into the momentum jump condition
at the interface \eqref{eq:jumpmomentum} leads  to
\begin{equation}
   (\tilde m -m\liq)u\liq^* +I\liq+  \dpi = (\tilde m-m\vap)u\vap^* +I\vap.
   \label{eq:hllp1}
\end{equation}
Further, we can eliminate the velocity $\sint$ from the mass jump condition at the interface \eqref{eq:jumpmass}:
\begin{equation}
   \frac{\tilde m}{\rho\liq^*}-u\liq^*=\frac{\tilde m}{\rho\vap^*}-u\vap^*,
   \label{eq:hllp2}
\end{equation}
and then insert the definition of the two outer mass fluxes \eqref{eq:ml} and \eqref{eq:mr} to obtain 
\begin{equation}
   u\liq^*\left(1-\frac{\tilde m}{m\liq} + \frac{\tilde m}{m\vap}S\liq\right) = u\vap^*\left(1-\frac{\tilde m}{m\vap} + \frac{\tilde m}{m\vap}S\vap\right).
   \label{eq:hllp3}
\end{equation}
The two equations \eqref{eq:hllp1} and \eqref{eq:hllp3} only include the two unknowns $u\liq^*$ and $u\vap^*$. All other quantities are known from the
initial data and the current iteration of $\tilde m$. Thus, we can  solve for the velocities in the inner states:
\begin{gather}
   u\vap^*=\frac{{(\tilde m - m\liq)(\frac{\tilde m}{m\vap}S\vap-\frac{\tilde m}{m\liq}S\liq)}/{(1-\frac{\tilde m }{m\liq})}+I\liq+\dpi -I\vap}{(\tilde m -
   m\vap)-(\frac{1-\tilde m /m\vap}{1-\tilde m /m\liq})(\tilde m- m\liq)},\label{eq:vstarvap}\\
   u\liq^*=\frac{{(\tilde m - m\vap)(\frac{\tilde m}{m\liq}S\liq-\frac{\tilde m}{m\vap}S\vap)}/{(1-\frac{\tilde m }{m\vap})}+I\vap -I\liq-\dpi}{(\tilde m -
   m\liq)-(\frac{1-\tilde m /m\liq}{1-\tilde m /m\vap})(\tilde m- m\vap)}.\label{eq:vstarliq}
\end{gather}
With the velocities in the inner states known, the pressures can be obtained from the outer momentum jump conditions \eqref{eq:il},\eqref{eq:ir}:
\begin{gather}
   p\vap^*=I\vap-m\vap u\vap^*,\\
   p\liq^*=I\liq-m\liq u\liq^*.
\end{gather}
Finally, the conservative inner states can be defined from the outer mass and energy jump conditions as
\begin{align}
   {\U}^*_i= \begin{pmatrix}
      \rho^*_i\\
      \rho^*_iu^*_i\\
      \rho^*_iv^*_i\\
      \rho^*_iw^*_i\\
      \rho^*_ie^*_i
   \end{pmatrix}
   =
   \begin{pmatrix}
      \frac{m_i}{u^*_i-S_i}\\
      \frac{m_i}{u^*_i-S_i}u^*_i\\
      \frac{m_i}{u^*_i-S_i}v_i\\
      \frac{m_i}{u^*_i-S_i}w_i\\
      \frac{E_i-p^*_iu^*_i }{u^*_{i}-S_i}
   \end{pmatrix},
   \label{eq:HLLPcons}
\end{align}
where the index $i$ refers to the liquid or vapor state $i\in[\mathrm{vap},\mathrm{liq}]$. Note that we apply the general HLLC methodology for the tangential
velocities and keep these untouched from the effects of the Riemann Problem. The velocity of the phase interface can be calculated from the mass jump  condition
\begin{equation}
   \sint=\frac{\rho\vap^* u\vap^*-\rho\liq^*u\liq^*}{\rho\vap^*-\rho\liq^*}.
   \label{eq:sihllp}
\end{equation}
With the inner states known, we can evaluate the evaporation model following \citet{Fechter2018} by using the conservative part of the inner extended
state vector only:
\begin{gather}
   \dot m     = \dot m ({\U}\liq^{*},{\U}\vap^{*}), \label{eq:mk} \\ 
   \dot q\vap = \dot q\vap ({\U}\liq^{*},{\U}\vap^{*}). 
\label{eq:qk} 
\end{gather}
Similar to the exact solver, we assume the vapor heat flux at the interface being identical to the one calculated from the evaporation model. The remaining
liquid heat flux can be calculated from the energy jump condition across the interface \eqref{eq:jumpenergy}:
\begin{equation} \dot q\liq=\tilde
   m(e\vap^{*}-e\liq^{*})+p\vap^{*} u\vap^{*}-p\liq^{*} u\liq^{*} + \dot q\vap -\dpi S^{I}. \label{eq:q} 
\end{equation}
Finally, we need to decide if our current guess for the interfacial mass flux $\tilde m$ is good enough by evaluating the
kinetic relation \eqref{eq:kineticrelation}. Based on the value of $\mathcal{K}$ a new guess is calculated with a Newton algorithm. This procedure is repeated
until the kinetic relation is fulfilled up to a user defined tolerance $\Delta_{\mathrm{HLLP}}$.

The numerical flux of the HLLP solver is calculated by applying the integral conservation principle across the approximate Riemann fan:
\begin{align}
   F^*\liq&=F(\U\liq) + S\liq({\U}\liq^*-\U\liq) - S^M {\U}\liq^* + (0,0,q\liq)^\mathrm{T},\label{eq:nfhllpliq}\\
   F^*\vap&=F(\U\vap) + S\vap({\U}\vap^*-\U\vap) - S^M {\U}\vap^* + (0,0,q\vap)^\mathrm{T},\label{eq:nfhllpvap}
\end{align}
where the two latter terms again refer to the ALE contribution and the interfacial heat fluxes, respectively.

\subsubsection{HLLP0 Approximate Riemann Solver}
The HLLP solver, described before, is a huge improvement compared to the exact Riemann solver with respect to computational effort. In this subsection, we introduce a further simplification, with which we can remove the iterative feature of the Riemann solver entirely. Therefore, we call it the
HLLP0 Riemann solver.

Consider again the approximate wave fan from Fig. \ref{fig:HLLP}. The need of an iteration in the HLLP solver stems from the fact of satisfying simultaneously 
the integral jump conditions \eqref{eq:jumps} and the kinetic relation \eqref{eq:kineticrelation}. Both sets of conditions are linked via the EOS making an
analytical solution of the problem impossible. In the context of the approximate Riemann solver, the simultaneous fulfillment is even dubious. We fulfill the
integral jump conditions for the extended and thermodynamically inconsistent state $\tilde \U^*$ but evaluate the evaporation model with the conservative states
$\U^*$. Therefore, jump condition and kinetic relation are never fulfilled by the same set of states. This becomes clear when looking at the pressure. Although
$\U^*\subset\tilde \U^*$, the pressures for each state definition will generally not coincide.

Given the realization that the tuple of states fulfilling the jump conditions \eqref{eq:jumps} and Eq. \eqref{eq:kineticrelation} are not identical, we can also 
separate them entirely. Therefore, we start the HLLP0 solver by evaluating the evaporation model not with the interfacial states but with the initial states:
\begin{gather}
   \dot m     = \dot m ({\U}\liq,{\U}\vap), \label{eq:mkhllp0} \\ 
   \dot q\vap = \dot q\vap ({\U}\liq,{\U}\vap). 
\label{eq:qkhllp0} 
\end{gather}
Similar to the guesses of the wave speeds \eqref{eq:wavespeeds}, which we keep, this gives us an estimate of the mass and vapor heat flux at the interface. From
there on, we simply apply Eqs. \eqref{eq:vstarvap}-\eqref{eq:sihllp}, as well as \eqref{eq:q} to calculate the inner states. The calculation of the numerical
fluxes is then done analogously to the HLLP solver.

In this approach, a kinetic relation is not needed since we define the interfacial fluxes with the initial states. We therefore control the entropy production
across the jump $\U\liq$, $\U\vap$ and not directly across the interface. At first, this seems counterproductive. However, referring to the discussion in the
beginning of this section, this does not differ from the HLLP solver since it is also not able to control the entropy production across the same set of states
as the integral jump condition. Furthermore, only the outer states are fully consistent with the EOS. Hence, it is more natural to use these states only to
evaluate the EOS and not a lumped version of the overdetermined inner states.

\section{Results}\label{sec:res}
In this section we apply the three proposed Riemann solvers to two representative test cases. First, we consider the evaporation shock tube discussed in
\citet{Hitz2020}. We compare our sharp-interface method with the molecular dynamics results of the authors. This is done with two different setups. First,
we use the ALE-method, so that only the numerical fluxes are used from the Riemann solver. Second, we use the Ghost Fluid method and allow the interface to sweep
across the mesh. Thus, we investigate the effects of setting the Ghost state as discussed in section \ref{sec:si}. Finally, we want to apply our Riemann solvers
to a complex, two-dimensional test case: a shock-droplet interaction.

\subsection{Evaporation Shock Tube - Moving Mesh}\label{sec:resmm}
We begin our numerical investigations by considering the two-phase evaporation shock tube cases studied in \citet{Hitz2020}. Therein, three different Riemann
problems for the Lennard-Jones truncated and shifted fluid (LJTS) were studied with molecular dynamics simulations and a sharp-interface method. A saturated
liquid was put in contact with a superheated vapor with three different amounts of superheat. In this work, we reconsider these test cases and apply our
sharp-interface method with the three different Riemann solvers. As EOS, we employed the PeTS EOS of \citet{Heier2018} and followed \citet{Homes2021} in using the
model for thermal conductivity of \citet{Lautenschlaeger2019} in the liquid region and of \citet{Lemmon2004} in the vapor.  The computational domain $x \in
[-200,1000]$ was discretized with $N_{Elems}=240$ elements. We chose a polynomial degree of the DG solution of $N=3$. In the bulk fluids, the HLLC Riemann solver
was used. We employ the above described ALE algorithm and move the mesh together with the phase interface. In this way, all sub-cells contain the same phase
during the calculation and the method is conservative up to machine accuracy. 

The initial data for the liquid and vapor states of the three test cases are summarized in Table \ref{tbl:initialdata}. When considering the
LJTS, all variables are non-dimensionalized, indicated with a bar. The reference values are $\sigma_{\mathrm{ref}}=1\,\si{\angstrom}$ as reference length,
$\epsilon_{\mathrm{ref}}/k_{\mathrm{B}}=1\,\si{\kelvin}$ reference energy and $m_{\mathrm{ref}}=1\,\si{\atomicmassunit}$ reference mass.  The reference time is given by
$t_{\mathrm{ref}}=\sigma_{\mathrm{ref}} \sqrt{m_{\mathrm{ref}}/\epsilon_{\mathrm{ref}}}$. A comprehensive summary of the non-dimensionalization can be found in
\citet{Merker2012}. 

\begin{table}
   \caption{Initial data for the evaporation shock tubes from \citet{Hitz2020}}
\centering
\begin{tabular}{l|c|c|c||c|c|c|}
& $\bar \rho_{\liq}$ & $\bar v_{\liq}$ & $\bar T_{\liq}$ & $\bar \rho_{\vap}$ & $\bar v_{\vap}$ & $\bar T_{\vap}$  \\
\hline
Case 1  & 0.6635 & 0 & 0.9 & 0.017800 & 0 & 0.8 \\ 
Case 2  & 0.6635 & 0 & 0.9 & 0.013844 & 0 & 0.8 \\ 
Case 3  & 0.6635 & 0 & 0.9 & 0.009889 & 0 & 0.8 \\ 
\end{tabular}
\label{tbl:initialdata}
\end{table}

\begin{figure}
   \centering
   \includegraphics[width=\linewidth]{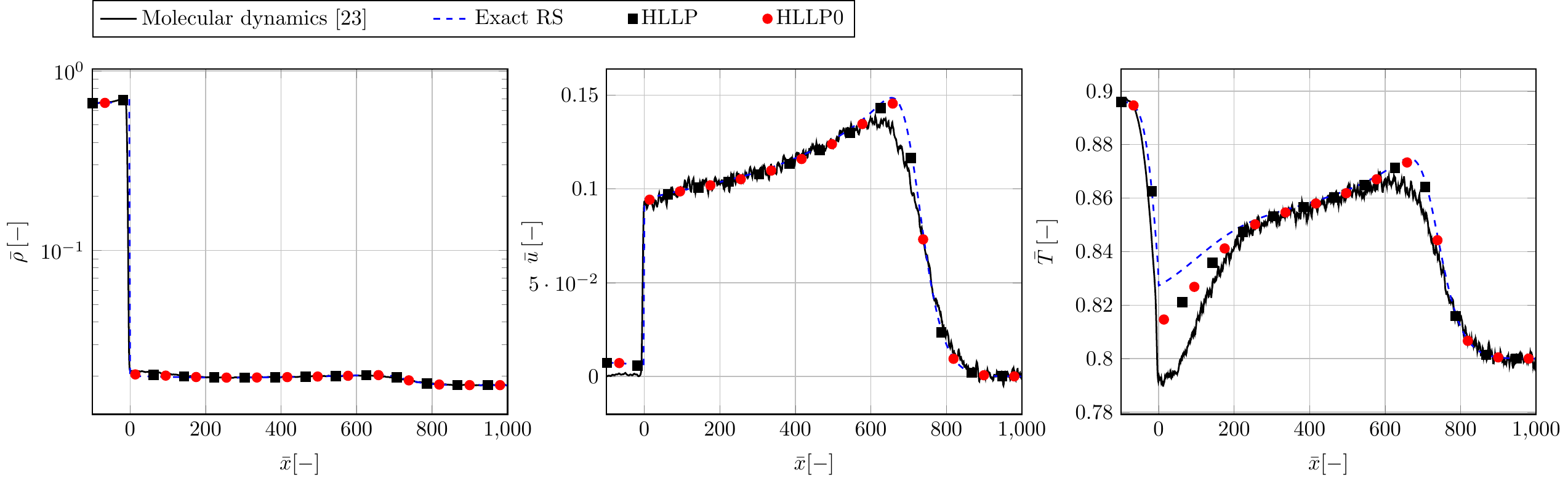}
   \caption{Case 1 at $\bar t=600$: Comparison between molecular dynamics from \cite{Hitz2020} and sharp-interface ALE method using three different Riemann solvers.}
   \label{fig:case1}
\end{figure}

Fig. \ref{fig:case1} shows the comparison of the molecular dynamics data from \cite{Hitz2020} with our sharp-interface method and the three proposed Riemann
solvers. For each solver, there is a very good agreement in the density, as well as in the velocity. Only directly at the shock wave, at about $x=700$, the sharp-interface method
predicts higher velocities than the molecular dynamics data. This aspect may be related to the absence of viscous terms in the mathematical fluid model while it is inherently contained in the molecular dynamics simulation. In the temperature
profile, the liquid side is in good agreement, as well as the vapor side in between shock and phase interface. At the shock wave, the temperature is 
slightly overpredicted, similar to the velocity profile. Near the interface, the vapor temperature is too high compared to the molecular dynamics data. However,
a qualitative agreement can be seen. We note that a small left-going rarefaction wave already left the computational domain and can therefore not be seen in
either results.

The
three Riemann solvers are in  excellent agreement with each other, except for the vapor temperature near the
interface. Here, both approximate Riemann solvers deviate from the exact solver and predict a slightly lower temperature. We want to emphasize, that the results,
presented here, are a strong improvement to the results shown in \citet{Hitz2020}. Therein, the interfacial vapor temperature was generally predicted higher than the
liquid interface temperature. This was in disagreement with the reference data and physical expectation. Here, the predicted solution always follows the
reference data qualitatively. Furthermore, the model of the Onsager coefficients by \citet{Cipolla1974} fully closes the Riemann solution. No fit of any coefficient is
needed to produce the results presented above, in contrast to the method from \cite{Hitz2020}.

\begin{figure}
   \centering
   \includegraphics[width=\linewidth]{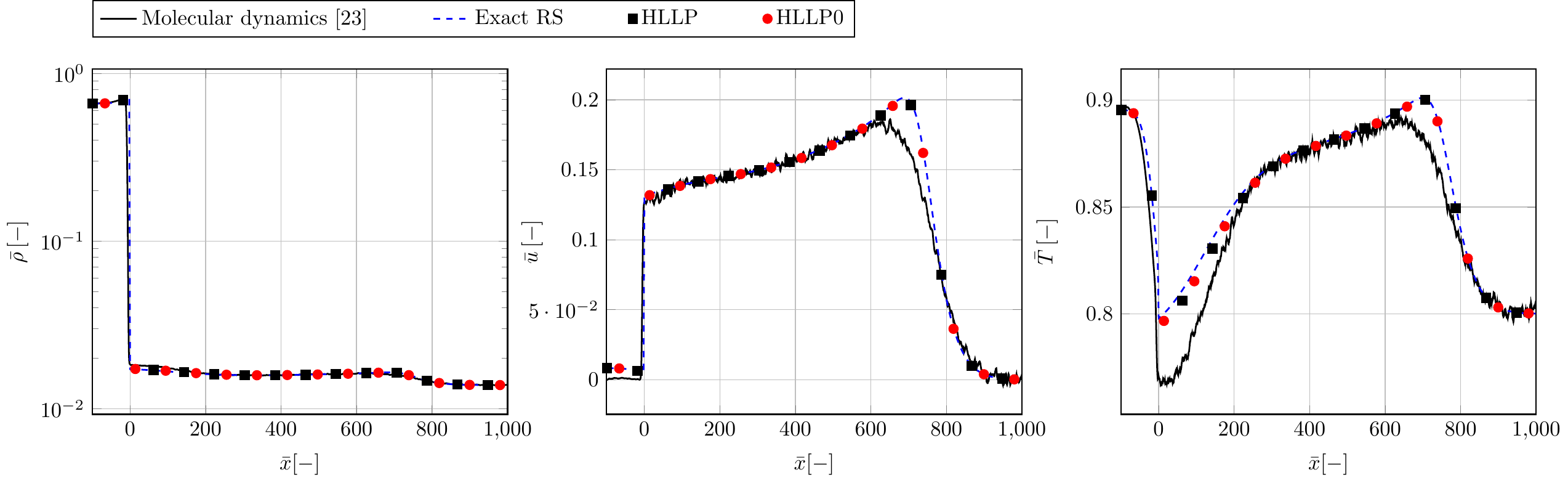}
   \caption{Case 2 at $\bar t=600$: Comparison between molecular dynamics from \cite{Hitz2020} and sharp-interface ALE method using three different Riemann solvers.}
   \label{fig:case2}
\end{figure}

\begin{figure}
   \centering
   \includegraphics[width=\linewidth]{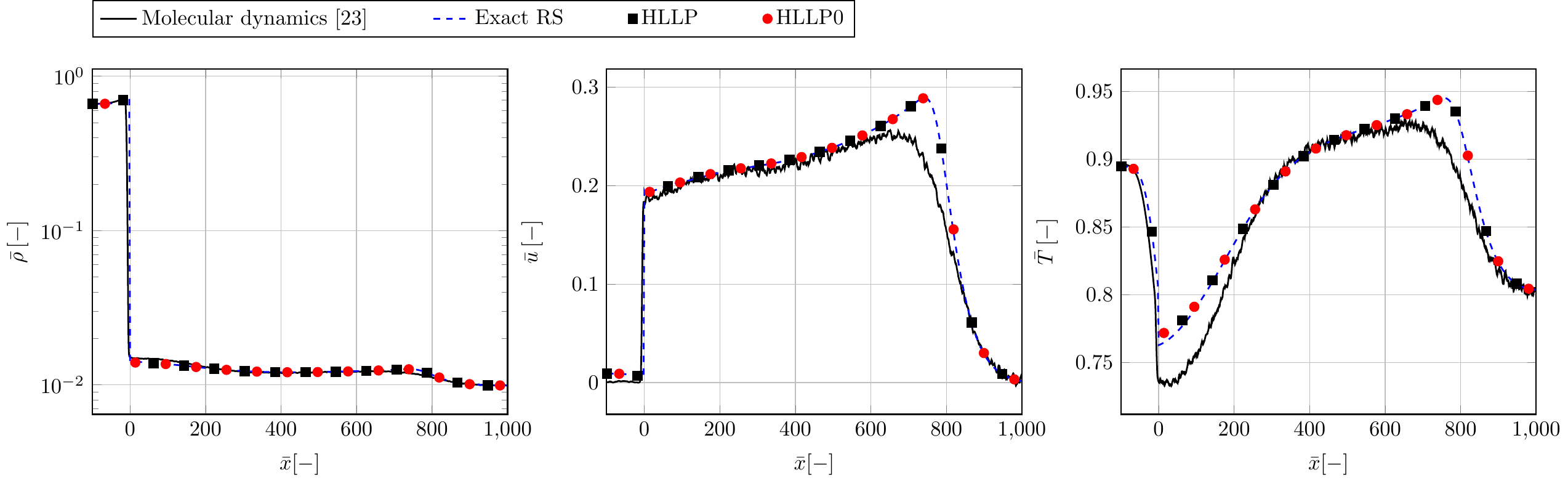}
   \caption{Case 3 at $\bar t=600$: Comparison between molecular dynamics from \cite{Hitz2020} of and sharp-interface ALE method using three different Riemann solvers.}
   \label{fig:case3}
\end{figure}

Fig. \ref{fig:case2} and \ref{fig:case3} show the same comparison for case 2 and case 3, respectively. In general, the same observations can be made as for case
1. However, the discrepancy in the vapor temperature at the interface between the exact Riemann solver and the approximate Riemann solvers is reduced for case 2
and slightly inverted for case 3. Here, the approximate solver slightly overpredicts the exact solver. We can summarize, that all three Riemann solvers are able to
calculate accurate numerical fluxes for the considered test cases. The predictions are very close to the molecular dynamics data and in any
case in a qualitative agreement. We want to note, that the remaining differences in the vapor temperature are most likely related to the closure relations for
the Onsager coefficients. Preliminary calculations indicate, that a further improvement of the closure leads to an even better agreement in the temperature
profiles. 

\subsection{Evaporation Shock Tube - Ghost Fluid Method} \label{sec:resgf}
Due to the utilization of the ALE method in the previous calculations, the interaction of the interface Riemann solvers with the solution is confined to determine 
numerical fluxes. In this subsection we show results with the ghost fluid approach. When the mesh is fixed, the interface may travel across the grid. As described in section \ref{sec:si}, this leads to the use
of the ghost states and inherent non-conservativity in the grid cells adjacent to the phase interface. In the following, we investigate the effect of this procedure onto
the numerical solution as it is directly affected by the employed Riemann solvers. We modify the evaporation shock tube case by  adding an initial
velocity of $\bar u= 0.2 $ in both phases as seen in Tbl. \ref{tbl:initialdataghost} so that several grid cells change their phase during the computation. To minimize interaction with the domain boundary, we extended the domain to
the right to $x \in [-200,1500]$ and discretize it with $N_{Elems}=300$ in order to keep the same spatial resolution as before.

\begin{table}
\caption{Initial data for the modified evaporation shock tubes. Data from  \citet{Hitz2020} is supplemented by an initial velocity of $\bar u$ in both phases.}
\centering
\begin{tabular}{l|c|c|c||c|c|c|}
& $\bar \rho_{\liq}$ & $\bar v_{\liq}$ & $\bar T_{\liq}$ & $\bar \rho_{\vap}$ & $\bar v_{\vap}$ & $\bar T_{\vap}$  \\
\hline
Case 1  & 0.6635 & 0.2 & 0.9 & 0.017800 & 0.2 & 0.8 \\ 
Case 2  & 0.6635 & 0.2 & 0.9 & 0.013844 & 0.2 & 0.8 \\ 
Case 3  & 0.6635 & 0.2 & 0.9 & 0.009889 & 0.2 & 0.8 \\ 
\end{tabular}
\label{tbl:initialdataghost}
\end{table}

\begin{figure}
   \centering
   \includegraphics[width=\linewidth]{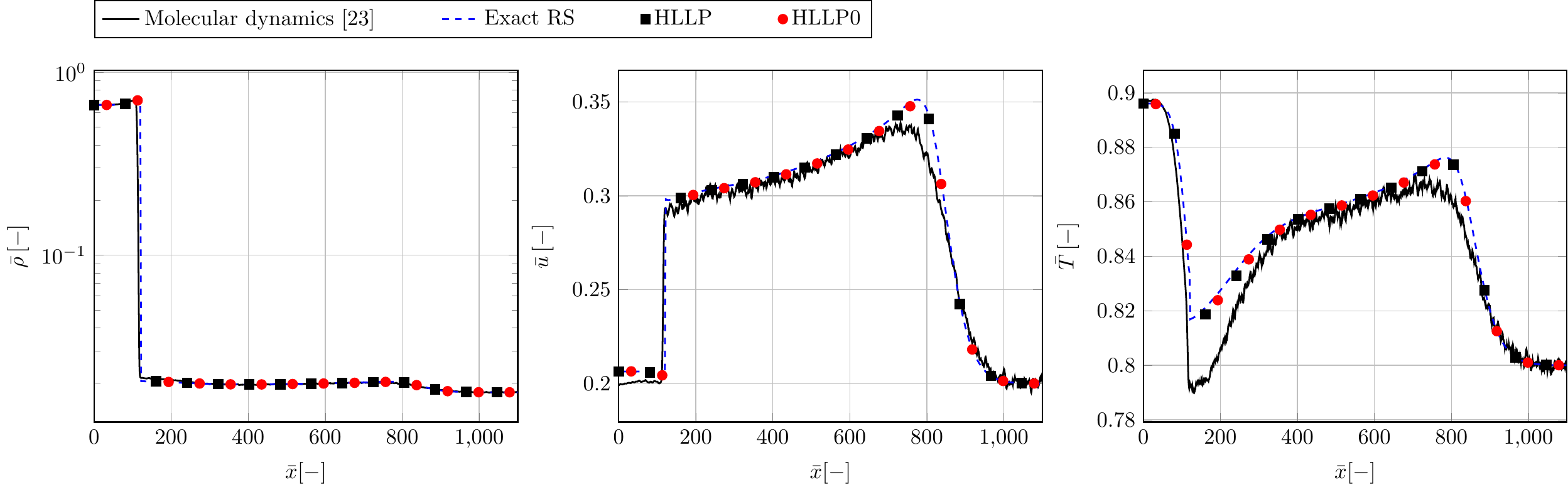}
   \caption{Modified case 1 at $\bar t=600$: Comparison between molecular dynamics from \cite{Hitz2020} and sharp-interface ghost fluid method using three different Riemann solvers.}
   \label{fig:case1ghost}
\end{figure}

Results for the modified case 1 from the sharp-interface Ghost Fluid method can be seen in Fig. \ref{fig:case1ghost}. We also included the molecular dynamics
data from \cite{Hitz2020} and translated it according to the superimposed velocity by $\bar u \bar t=120$. For both density and velocity profile, the
observations from the moving mesh cases still hold. In the temperature profile, the sharp-interface method generally predicts a higher temperature than the
molecular dynamics data, also away from the interface. In addition, the discrepancies between exact and approximate Riemann solvers has nearly vanished. The most
obvious difference is seen in the solutions with the exact Riemann solver: the vapor temperature at the interface is clearly lower compared to the moving mesh
case. 

\begin{figure}
   \centering
   \includegraphics[width=\linewidth]{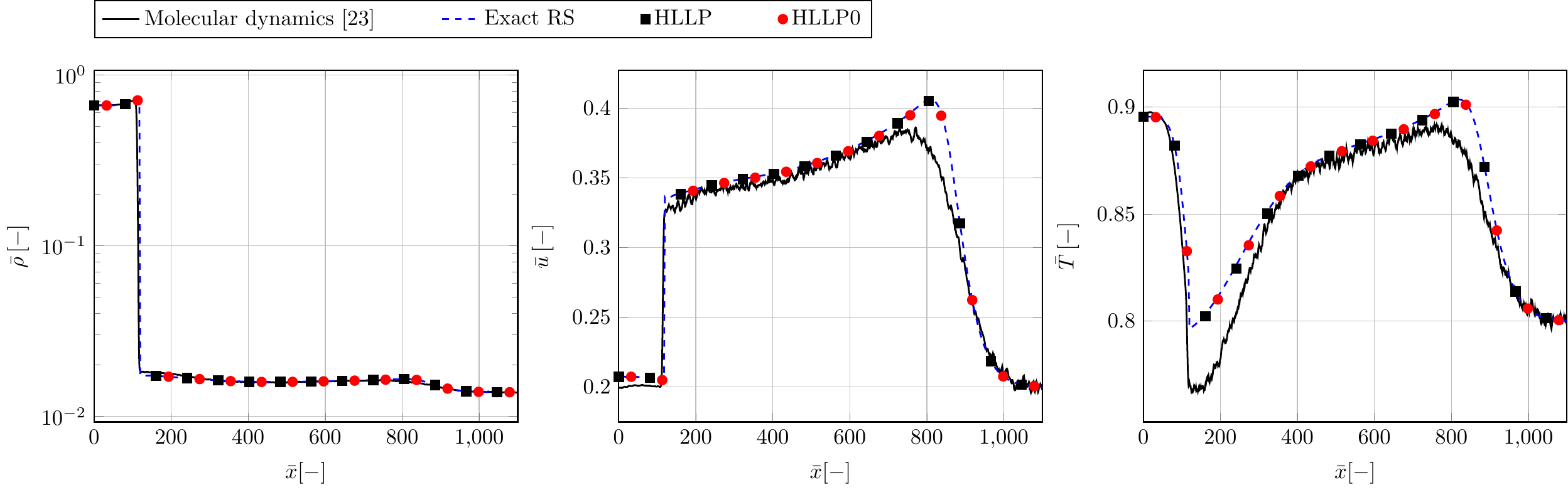}
   \caption{Modified case 2 at $\bar t=600$: Comparison between molecular dynamics from \cite{Hitz2020} and sharp-interface ghost-fluid method using three different Riemann solvers.}
   \label{fig:case2ghost}
\end{figure}

\begin{figure}
   \centering
   \includegraphics[width=\linewidth]{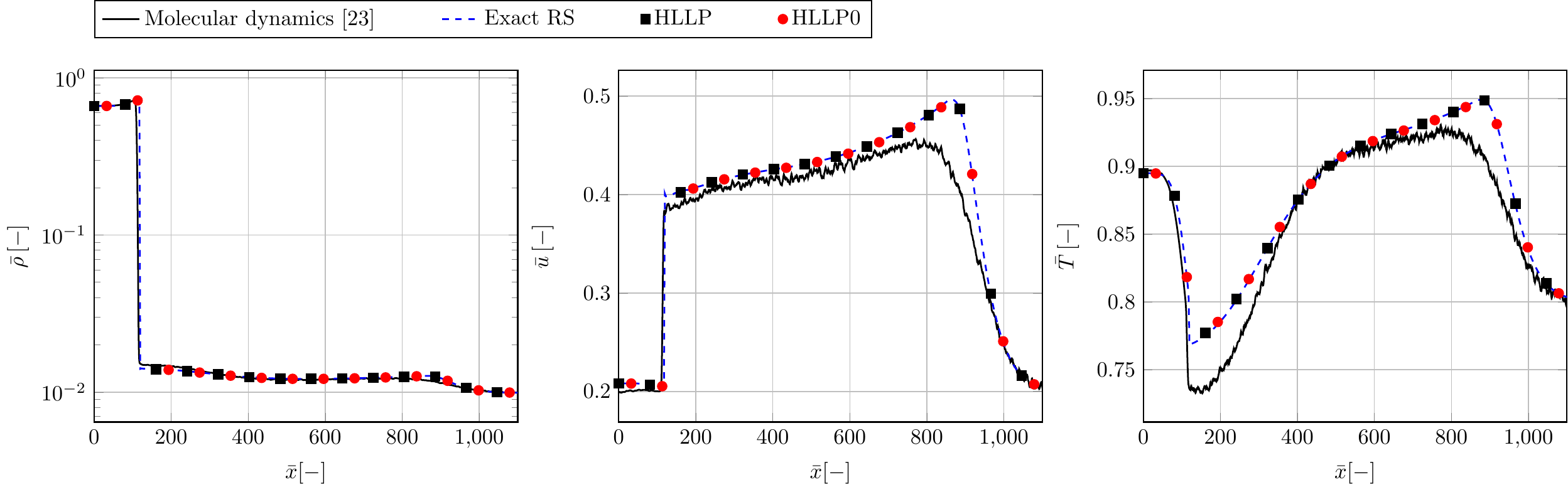}
   \caption{Modified case 3 at $\bar t=600$: Comparison between molecular dynamics from \cite{Hitz2020} of and sharp-interface ghost-fluid method using three different Riemann solvers.}
   \label{fig:case3ghost}
\end{figure}

Overall, the use of the Ghost Fluid method in contrast to the ALE method only slightly alters the numerical result. The biggest difference favors our numerical
method, as differences between exact and approximate Riemann solvers vanish. All three proposed Riemann solvers are clearly able to produce numerical fluxes and
corresponding ghost states which lead to numerical predictions in line with the molecular dynamics data. When considering the test cases 2 and 3 shown in Fig.
\ref{fig:case2ghost} and \ref{fig:case3ghost}, respectively, the same observations can be made. However, the general, but slight,  overprediction of the
temperature becomes more obvious. In addition, the same effect can be seen in the velocity profiles. The overall results of the ghost-fluid method are
nevertheless in a very good agreement with the molecular dynamics data.  

\begin{figure}
   \centering
   \includegraphics[width=\linewidth]{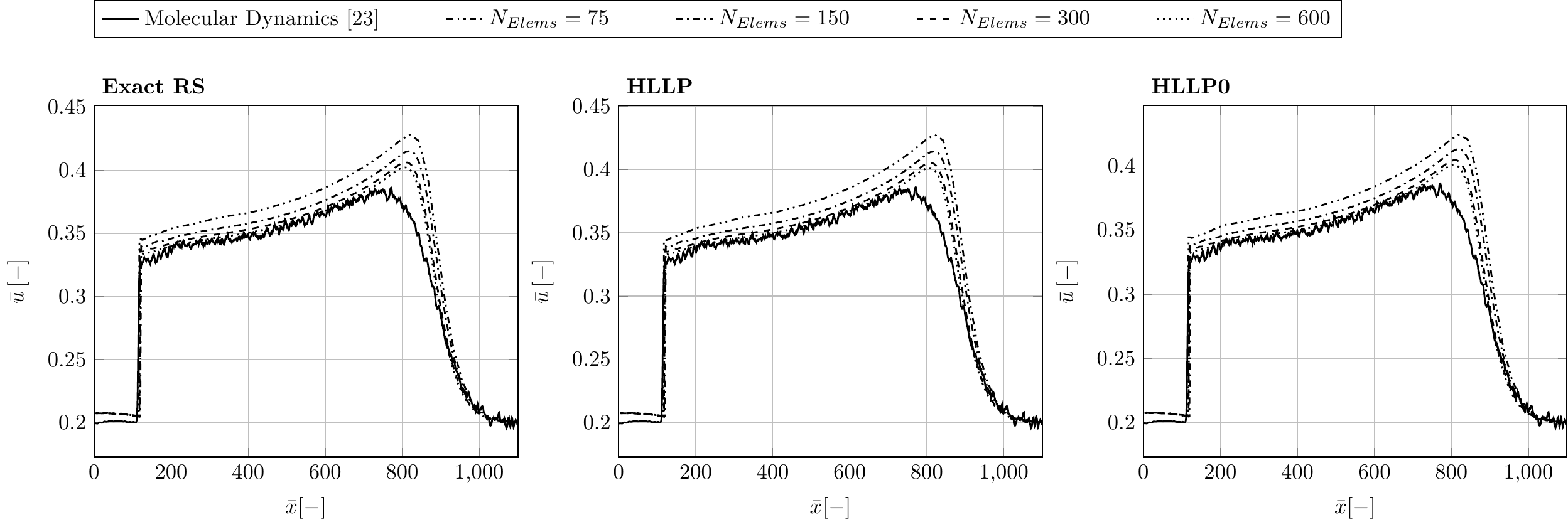}
   \caption{Modified case 2 at $\bar t=600$: Comparison between the velocity profile of molecular dynamics data from \cite{Hitz2020} of the present sharp-interface
   ghost-fluid method with three different mesh resolutions for the three different Riemann solvers. } \label{fig:gridu}
\end{figure}

\begin{figure}
   \centering
   \includegraphics[width=\linewidth]{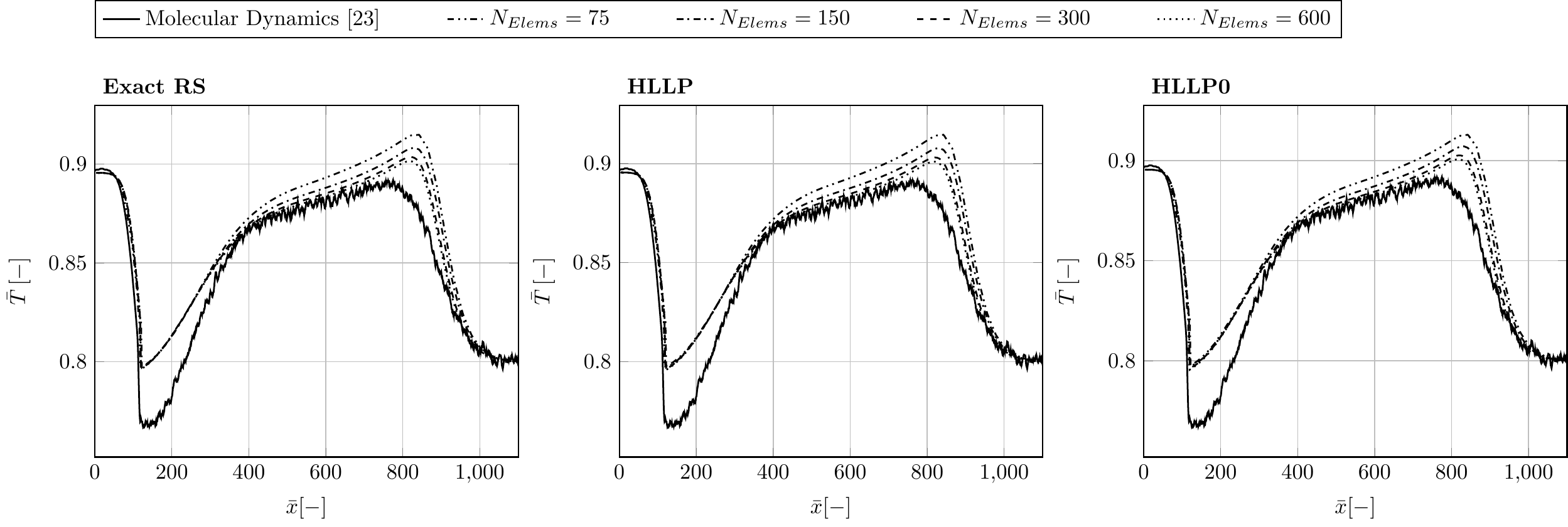}
   \caption{Modified case 2 at $\bar t=600$: Comparison between the temperature profile of molecular dynamics data from \cite{Hitz2020} of the present sharp-interface
   ghost-fluid method with three different mesh resolutions for the three different Riemann solvers. } \label{fig:gridt}
\end{figure}

Potential artifacts of the ghost-fluid method depend on the mesh size. The larger the grid elements, the more mass, momentum and energy are
stored in an individual cell and the conservation errors may get bigger. Therefore, we want to investigate how the grid size effects the numerical results. We
choose the modified test case 2 and run the simulation with grids of the size $N_{Elems}=75$, $N_{Elems}=150$, $N_{Elems}=300$ and $N_{Elems}=600$ for each
of the proposed Riemann solvers. Results for the velocity profile can be seen in Fig. \ref{fig:gridu}, whereas the temperature profile is shown in Fig.
\ref{fig:gridt}. For a decreasing grid size, the solution shows an increase in the vapor velocity and temperature. This indicates an overprediction in the
evaporation mass flux. On the other hand, the temperature in the liquid and in the vapor near the interface show no sign of grid sensitivity. All three Riemann
solvers have the same convergence behavior. Therefore, reduced grid sizes show no strong negative effects in the solution. However, in heavily underresolved cases,
one can expect an overprediction of the interfacial fluxes.

\subsection{Shock-Droplet Interaction}\label{sec:SD}
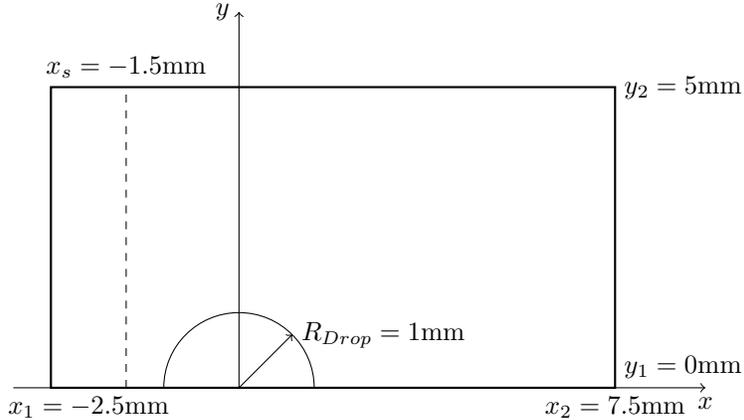
\begin{figure}
   \centering
   \begin{tikzpicture}
      \draw [->] (-3,0)--(6.2,0) node[below] {$x$};
      \draw [->] (-0,0)--(0,5) node[left] {$y$};

      \draw[thick]    (-2.5,0) rectangle (5,4);
      \node[below] at (-2,0) {$x_1=-2.5\si{\milli\meter}$};
      \node[below] at (5,0)  {$x_2= 7.5\si{\milli\meter}$};
      \node[right] at (5,4)  {$y_2=   5\si{\milli\meter}$};
      \node[above right] at (5,0)  {$y_1= 0\si{\milli\meter}$};
      \draw (-1.0,0) -- (1.0,0) arc(0:180:1.0) --cycle;
      \draw [->] (0,0)--(0.707,0.707) node[right] {$R_{Drop}=1 \si{\milli\meter}$};

      \draw[dashed] (-1.5,0)--(-1.5,4) node[above] {$x_s=-1.5\si{\milli\meter}$};
   \end{tikzpicture}
   \caption{Geometrical setup of the shock droplet interaction.}
   \label{fig:sd}
\end{figure}

As a final test case, we apply our proposed Riemann solvers to a complex, two-dimensional test case: a shock droplet interaction. We adopt the case of
\citet{Fechter2018}, in which this was already studied in the context of phase transition. In their work, the authors considered an initially resting, but
evaporating n-dodecane droplet, on which a shock wave impinges with a Mach number of $\mathrm{Ma}=1.5$. The geometrical setup is shown in Fig. \ref{fig:sd}. The droplet,
initially at rest, had a density of $\rho=539.94\si{\kilogram\per\cubic\meter}$ and a pressure of $p=1.3\si{bar}$. It is surrounded by quiescent vapor with a
density of $\rho=4.383\si{\kilogram\per\cubic\meter}$ and a pressure of $1 \si{\bar}$. A shock wave was positioned at $x_s=-1.5\si{\milli\meter}$ with a
post-shock density of $\rho=9.696\si{\kilogram\per\cubic\meter}$, velocity of $u=108.87\si{\meter\per\second}$ and pressure of $p=2.27\si{\bar}$. The initial
conditions are slightly altered to the ones used in \cite{Fechter2018}, since we chose a Peng-Robinson EOS to model the fluid. The relevant EOS parameters are
summarized in Tbl.  \ref{tbl:ndodecane}, where the subscript $c$ indicates critical values, $M$ the molar mass and $\omega$ the acentric factor. For the thermal
conductivity, the model of \citet{Chung1988} was used. Surface tension has been set to a constant value of $\sigma=9\si{\milli\newton\per\meter}$.

The computational domain was discretized with $512$ elements in $x$ direction and $256$ elements in $y$ direction. We only simulated a half droplet and
therefore prescribed symmetry boundary conditions at the bottom. Further, we used slip-wall boundary conditions at the top border, inflow boundary
conditions defined by the initial state at the left boundary and a supersonic outflow condition at the right boundary. The polynomial degree of the solution was
set to $N=3$ totaling the number of DoFs to roughly $2$ million. As Riemann solver in the bulk phases, the local Rusanov flux was chosen. 

\begin{table}
   \caption{EOS parameters of  the Peng-Robinson EOS for n-Dodecane.}
\centering
\begin{tabular}{c|c|c|c|c}
   $\rho_c \, [\si{\kilogram\per\cubic\meter}]$ & $p_c \,[\si{\bar}]$ & $ T_c\, [\si{\kelvin}]$ & $M\, [\si{\kilogram\per\mol}]$ & $\omega \, [-]$  \\
\hline
226.55 & 18.17 & 658.1 & 0.1703 & 0.576 \\
\end{tabular}
\label{tbl:ndodecane}
\end{table}

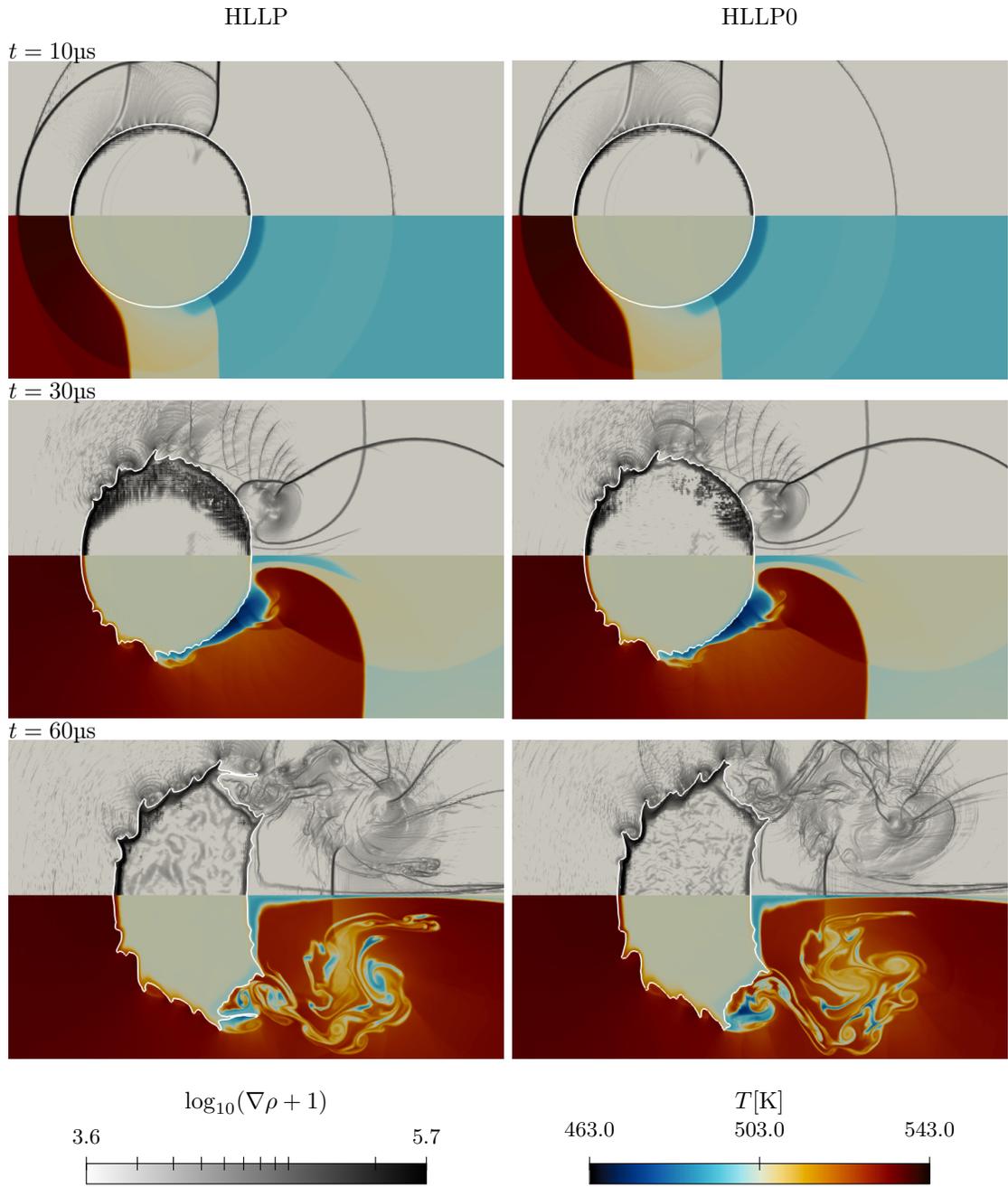
\begin{figure}
	\begin{tikzpicture}
      \node at (0,10.0){\includegraphics[trim=300 200 600 200,clip, width=.6\linewidth]{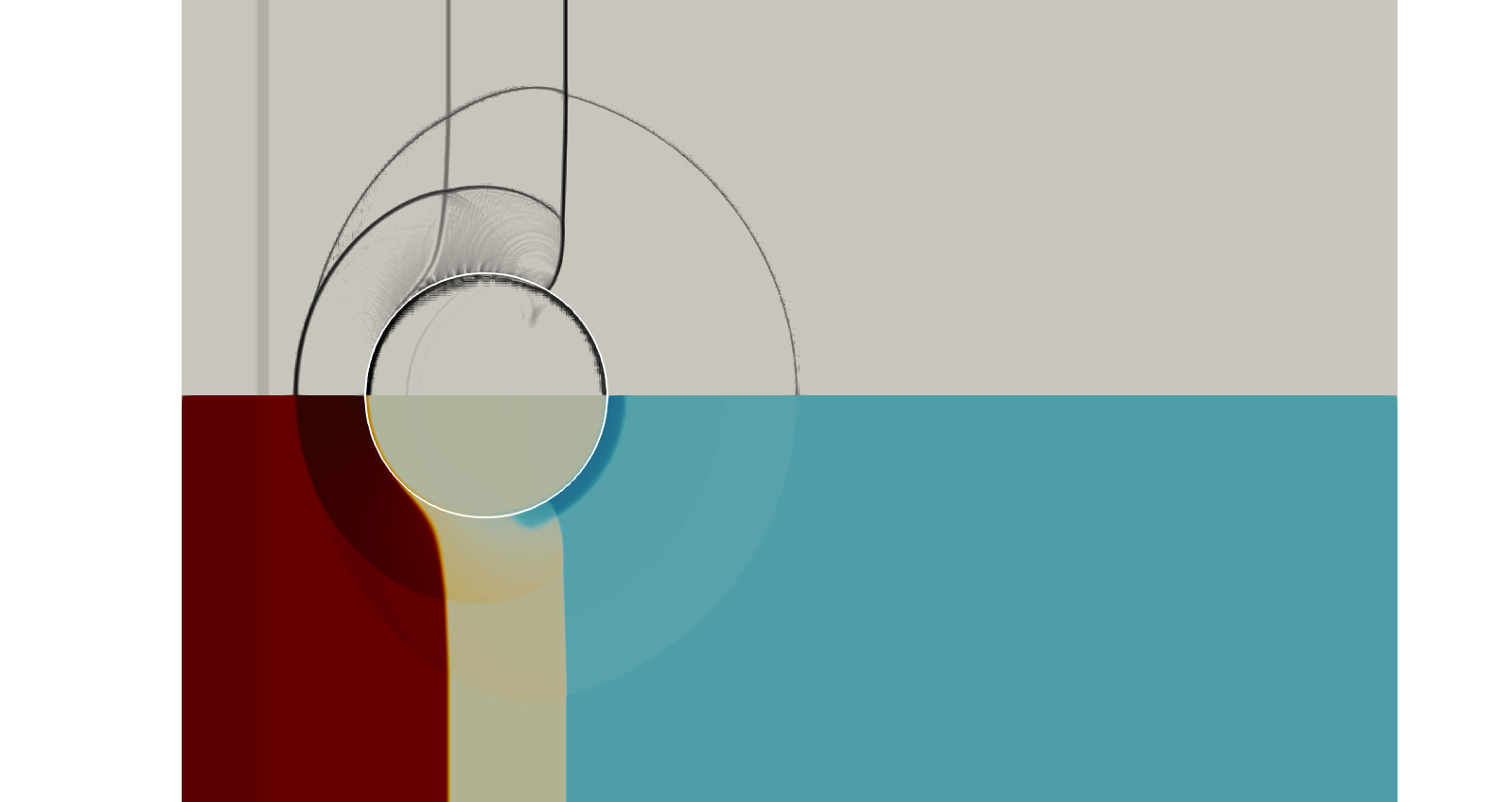}};
      \node at (0,05.0){\includegraphics[trim=300 200 600 200,clip, width=.6\linewidth]{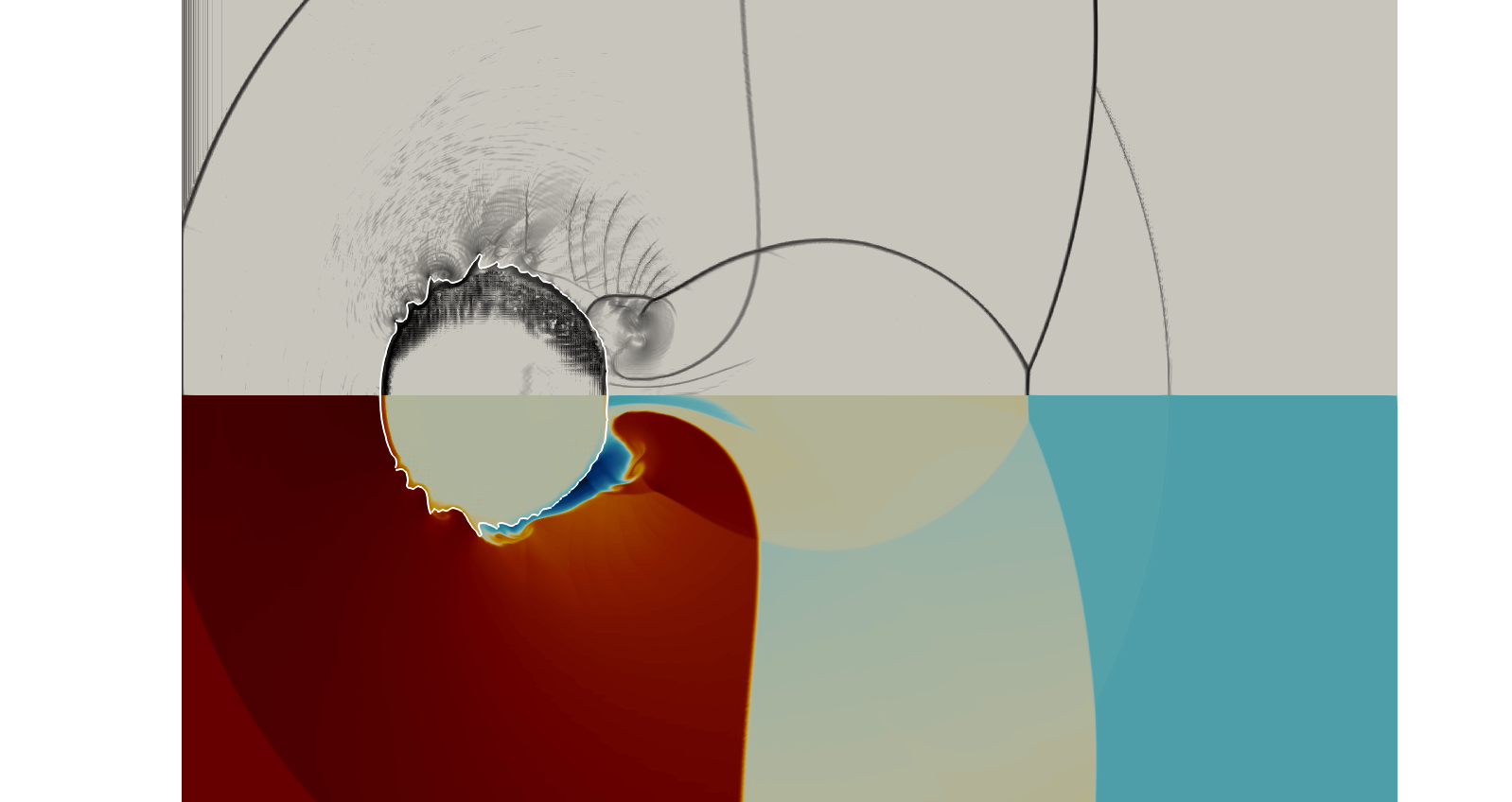}};
      \node at (0,00)  {\includegraphics[trim=300 200 600 200,clip, width=.6\linewidth]{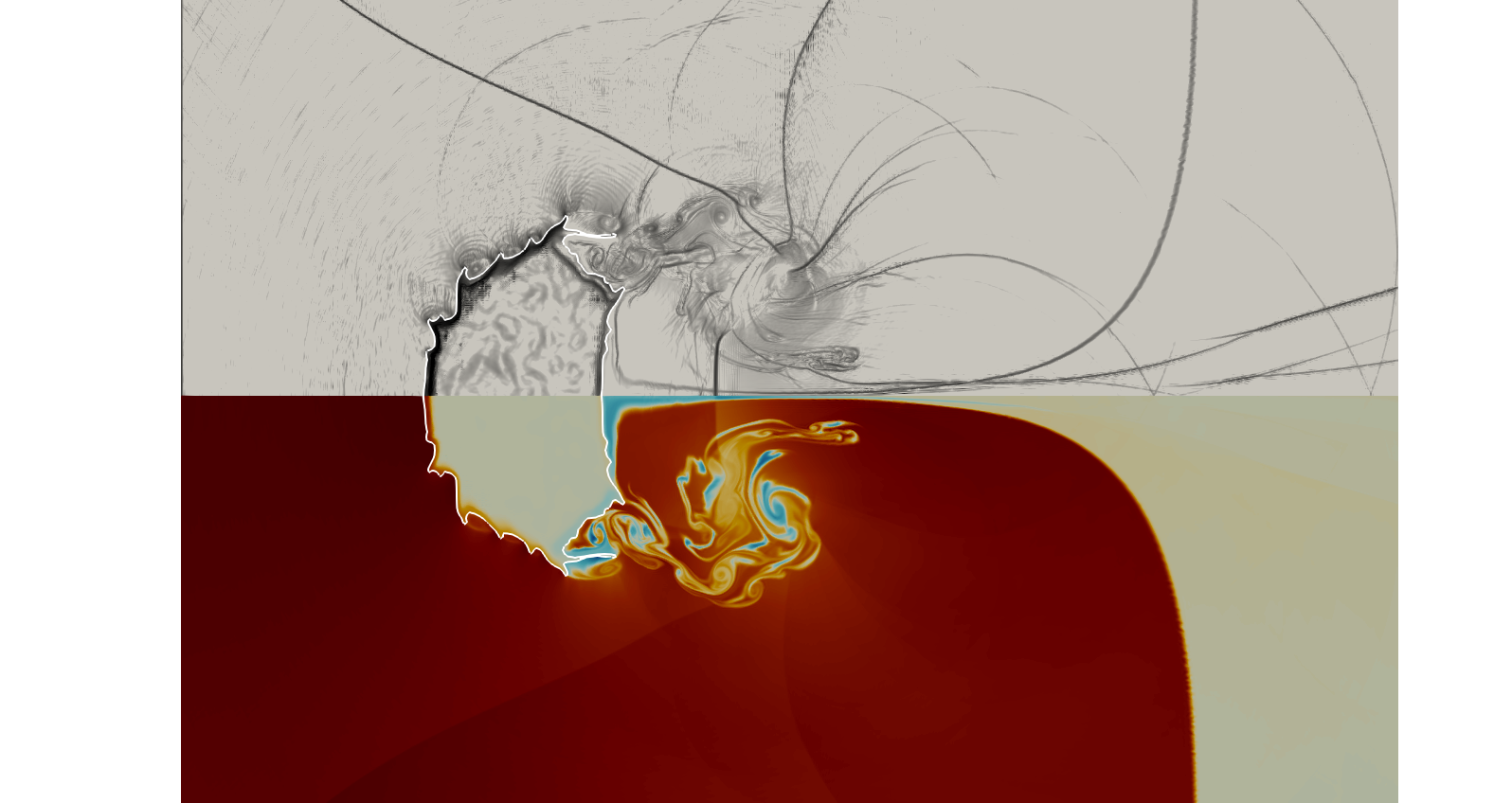}};
      \node at (7.4,10){\includegraphics[trim=300 200 600 200,clip, width=.6\linewidth]{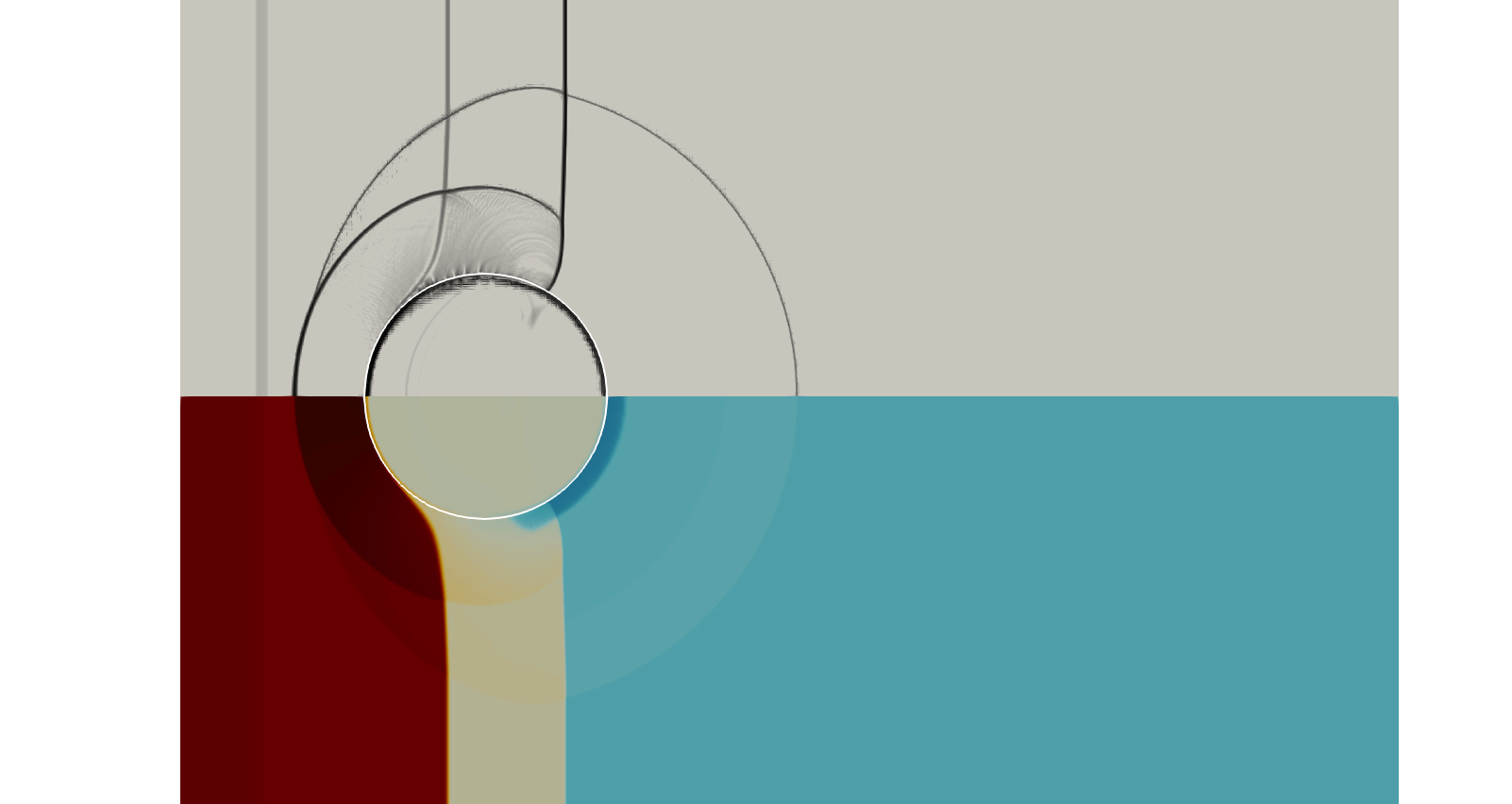}};
      \node at (7.4,05){\includegraphics[trim=300 200 600 200,clip, width=.6\linewidth]{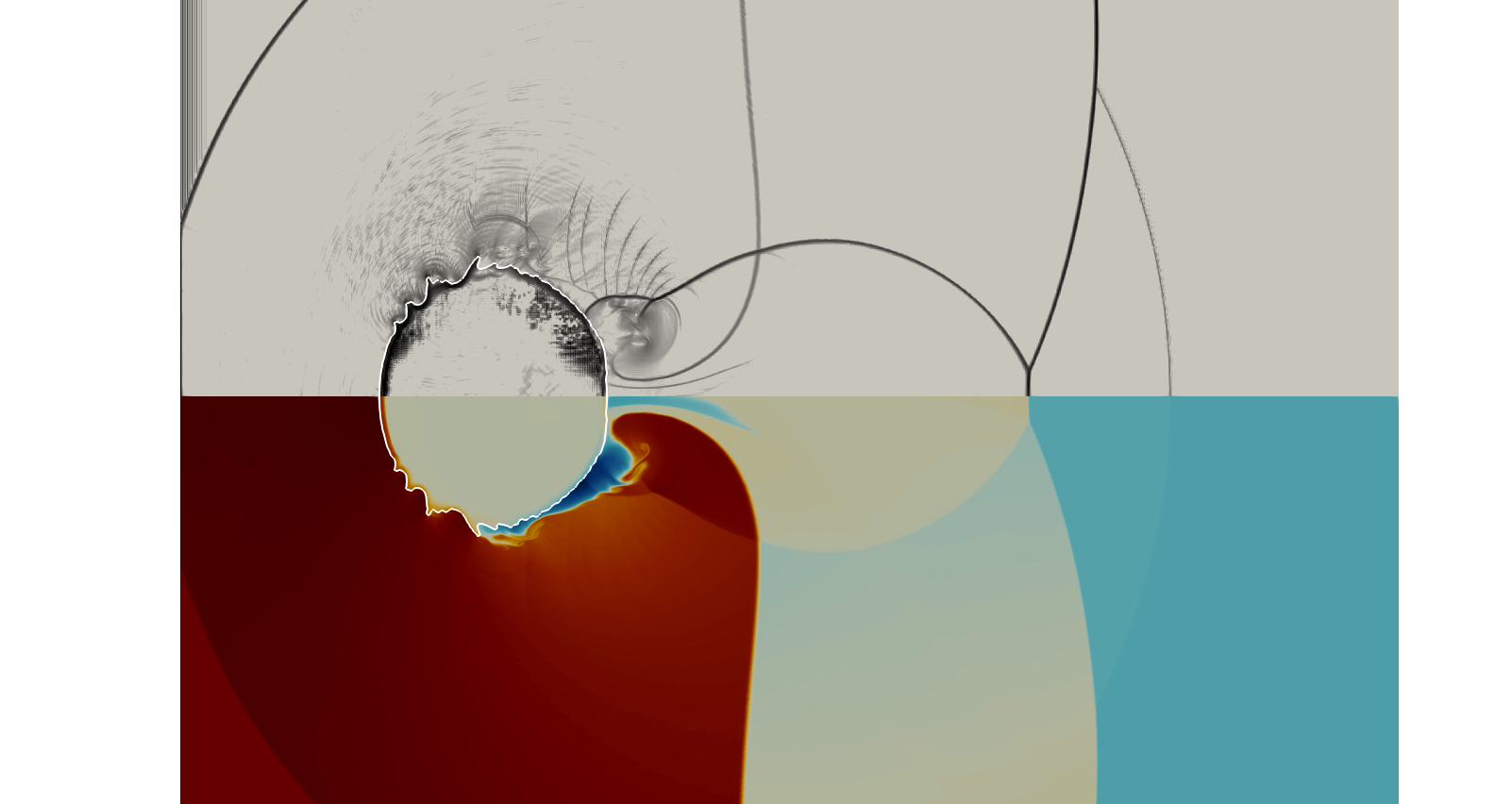}};
      \node at (7.4,00){\includegraphics[trim=300 200 600 200,clip, width=.6\linewidth]{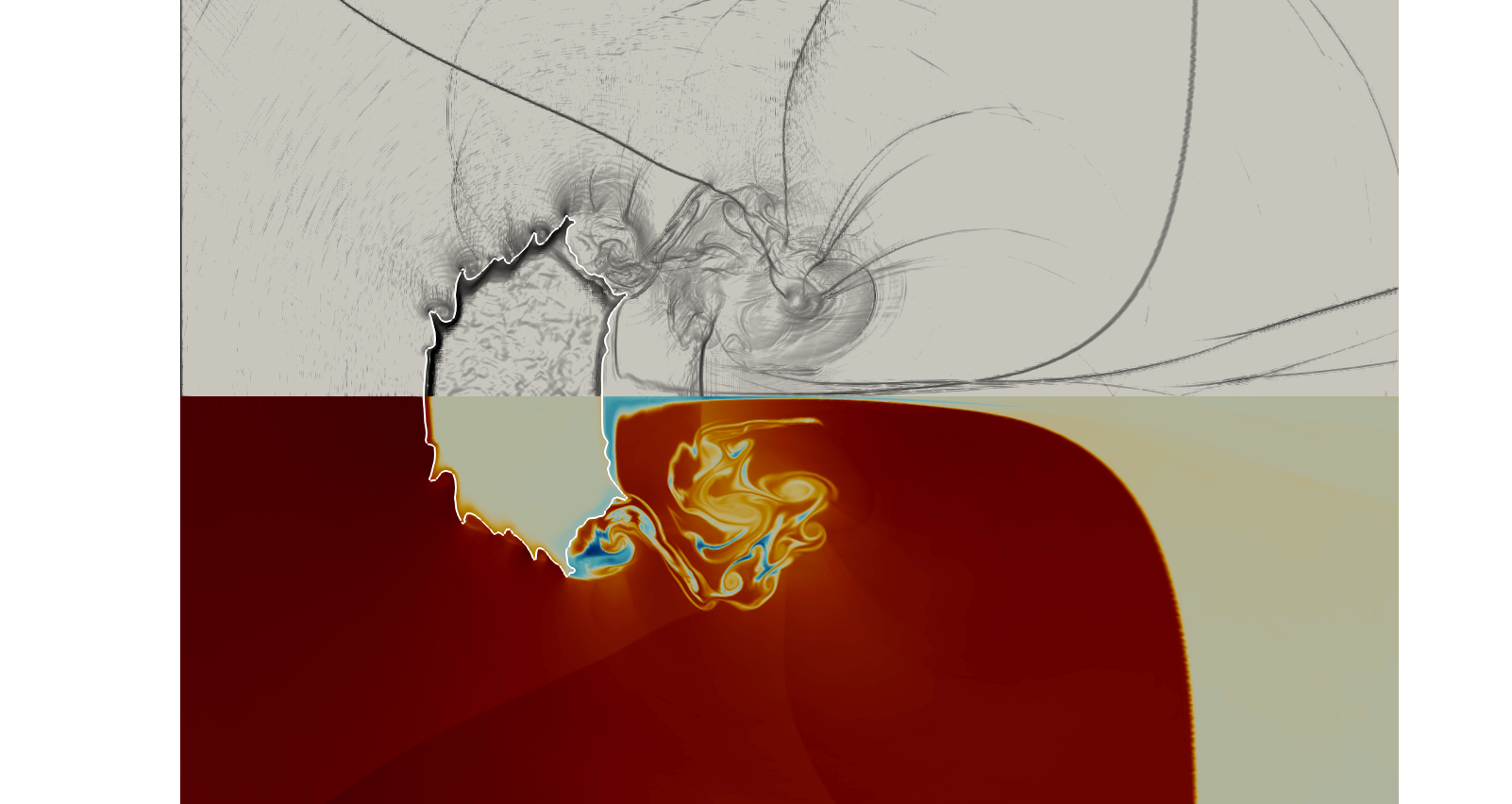}};

      \node at (0, -3.5)[scale=1]{\includegraphics{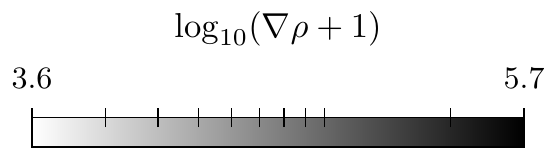}};
      \node at (7.4, -3.5)[scale=1]{\includegraphics{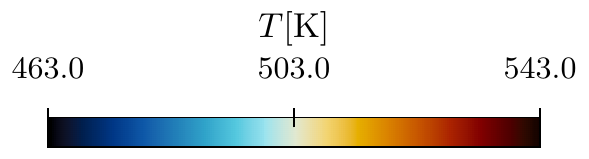}};
      
      \node at (-0.0,13.00)[scale=1] {HLLP};
      \node at ( 7.4,13.00)[scale=1] {HLLP0};
	
      \node at (-3.0,12.45)[scale=1] {$t=10\si{\micro\second}$};
      \node at (-3.0, 7.45)[scale=1] {$t=30\si{\micro\second}$};
      \node at (-3.0, 2.45)[scale=1] {$t=60\si{\micro\second}$};
	\end{tikzpicture}
   \caption{N-dodecane shock-droplet interaction calculated with the HLLP (left) and HLLP0 (right) Riemann solver. The interface is depicted as white line. The top half
   shows a numerical schlieren image, whereas the lower half shows the temperature field.}
   \label{fig:shockdrop}
\end{figure}

Because of the immense computational effort for the exact interface Riemann solver we only compare solutions for the two
approximate Riemann solvers with each other. Fig. \ref{fig:shockdrop} shows the numerical results after $t=10\si{\micro\second}$, $t=30\si{\micro\second}$
and $t=60\si{\micro\second}$. The top half of each time instance depicts a numerical schlieren image which shows areas of high density gradients in darker color.
The bottom half visualizes the temperature field, where red and blue areas indicate higher and lower temperatures compared to the initial droplet temperature,
respectively. After $t=10\si{\micro\second}$, the impinging shock wave has passed roughly two thirds of the droplet surface. It is followed by a contact
discontinuity due to the imperfect numerical initialization. The droplet was initially evaporating, as can be seen by the low vapor temperature around the
droplet at the areas the shock has not reached yet. Further, a weaker shock wave is visible, clearly preceding the impinging shock wave. This wave was generated
across the full droplet surface and is therefore also seen in front of the droplet, interacting with the bow shock. On the front half of the droplet, a slight
temperature increase in the liquid can be seen. Here, the initially evaporating surface now condensates. A similar observation was made by \citet{Fechter2018}.
The hot, accelerated vapor impinges on the cooler droplet surface and changes its phase. Thus, latent heat is released leading to a heating of the liquid droplet
surface. At this early stage, no differences between the two interface Riemann solvers can be seen.

After $t=30\si{\micro\second}$, both shock wave and contact discontinuity have fully passed the droplet. The high speed vapor flow around the droplet initiated
the characteristic instabilities on its surface. The whole droplet is slightly quenched and of an ellipsoidal shape. The effects of the condensing vapor on the
front of the droplet have become stronger as the surface has become hotter. On the backside, a region of cold vapor is attached to the droplet. It is initially
generated by the expansion waves induced in this area. The drop in temperature and pressure on the vapor side then lead to evaporation at the interface and
therefore cold vapor exiting the surface. The only difference between both Riemann solvers can be seen inside the liquid droplet where density variations occur
in a larger region for the HLLP solver.

Finally, after $t=60\si{\micro\second}$, the droplet has been totally deformed. The cold vapor, formerly attached to the back side of the droplet, is now
detached from the surface and dissipated in the wake. Still, effects of evaporation and condensation are clearly visible on the droplet surface, however the
core temperature of the droplet is unchanged. The difference between the two Riemann solvers is small. The HLLP solver leads to a slightly finer
structured droplet surface with the most prominent feature being the small ligament forming at the top. However, the general droplet
shape is identical for both Riemann solvers. The small differences in the surface structure has an effect on the temperature distribution in the wake but not on
its spatial extension. Given the results of section \ref{sec:resgf}, the deviations between both approximate solvers are probably related to strongly
underresolved regions on the droplet surface. These become more dominant in the solution as the droplet breaks up and small scale structures develop.
Nevertheless, both solvers predict a very similar flow solution and droplet surface for this complex, two-dimensional test case. 

\begin{figure}
	\begin{tikzpicture}
      \node at (0,10.0){\includegraphics[trim=300 200 600 200,clip, width=.6\linewidth]{HLLP_10_2.png}};
      \node at (0,05.0){\includegraphics[trim=300 200 600 200,clip, width=.6\linewidth]{HLLP_30_2.png}};
      \node at (0,00)  {\includegraphics[trim=300 200 600 200,clip, width=.6\linewidth]{HLLP_60_2.png}};
      \node at (7.4,10){\includegraphics[trim=870 272.4 1300 272.4 ,clip, width=.6\linewidth]{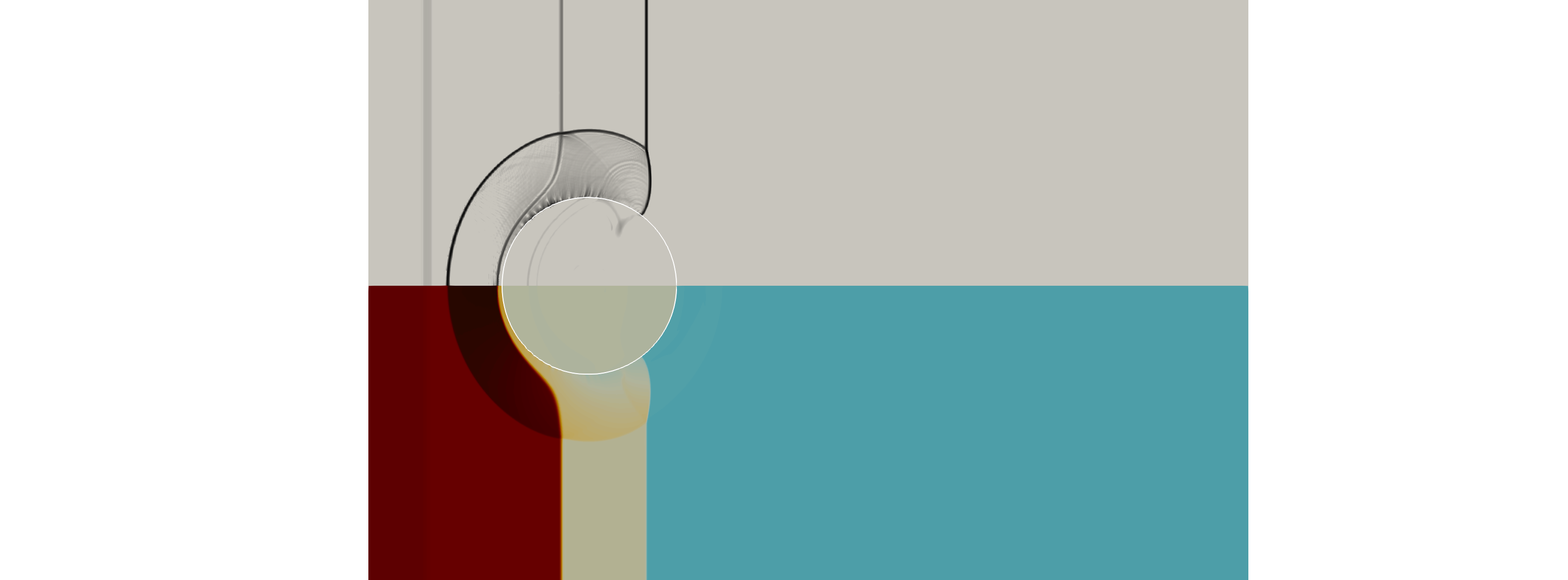}};
      \node at (7.4,05){\includegraphics[trim=870 272.4 1300 272.4 ,clip, width=.6\linewidth]{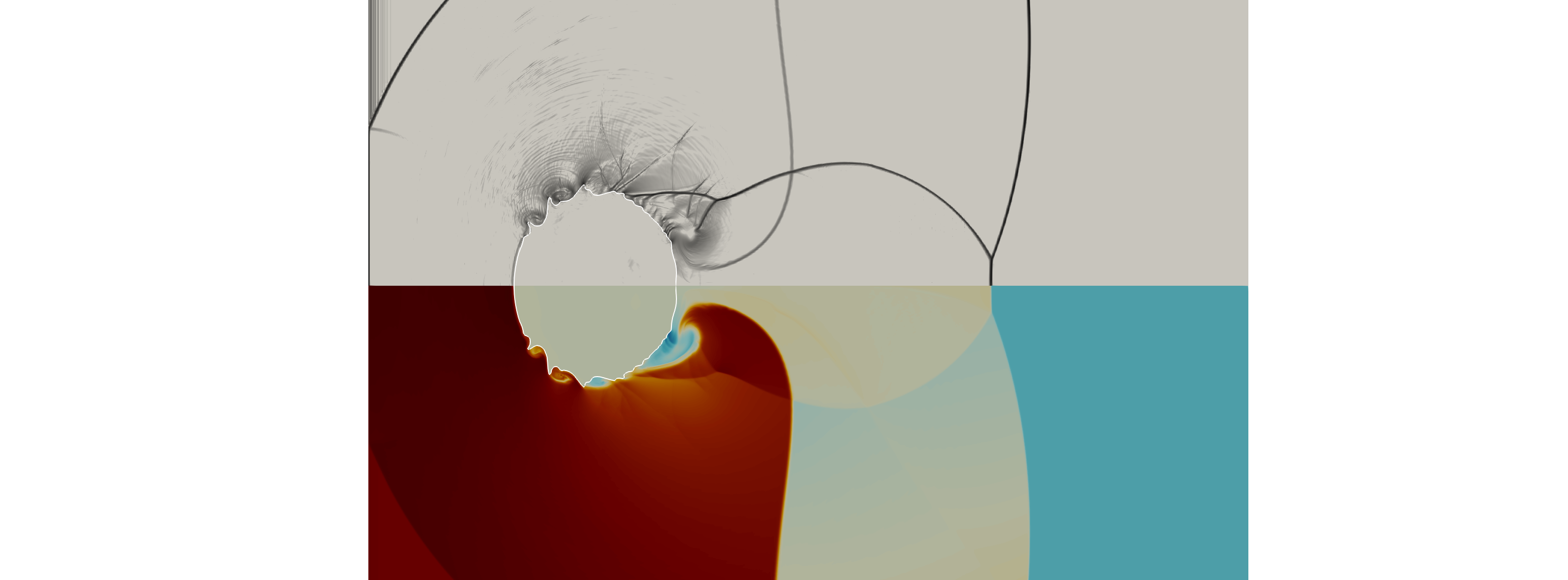}};
      \node at (7.4,00){\includegraphics[trim=870 272.4 1300 272.4,clip, width=.6\linewidth]{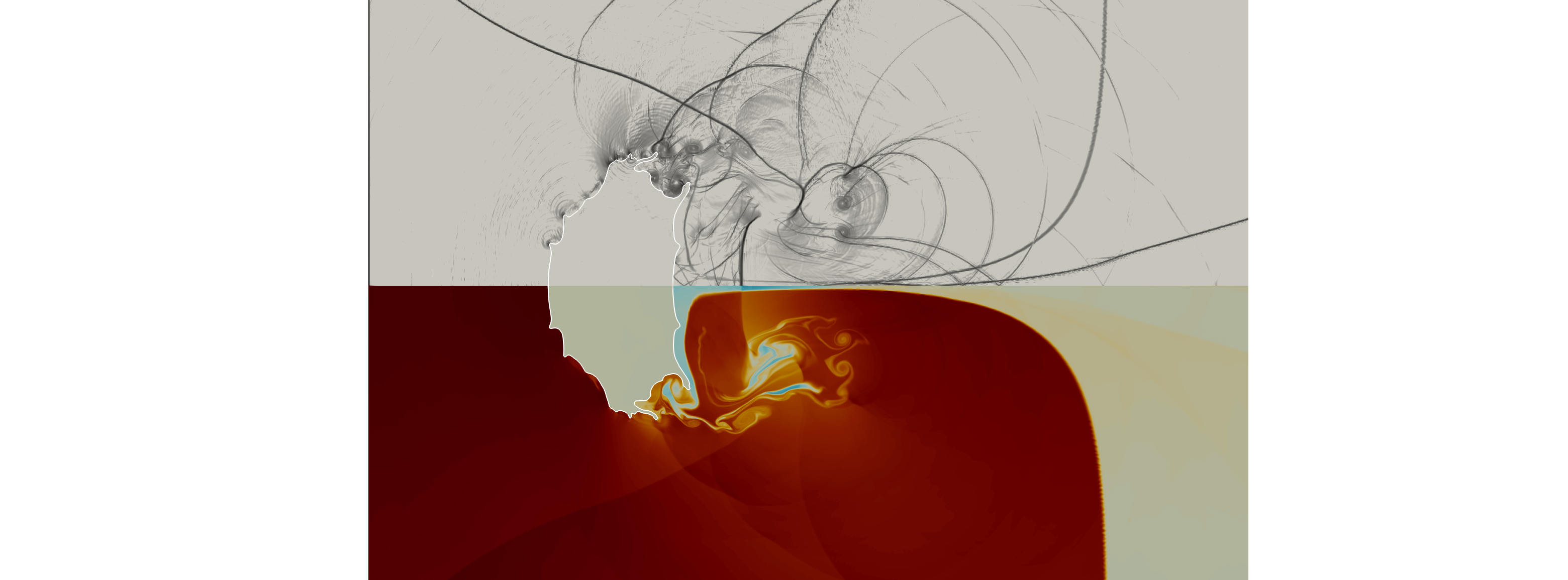}};

      \node at (0, -3.5)[scale=1]{\includegraphics{schlieren.pdf}};
      \node at (7.4, -3.5)[scale=1]{\includegraphics{T.pdf}};
      
      \node at (-0.0,13.00)[scale=1] {HLLP};
      \node at ( 7.4,13.00)[scale=1] {HLLC};
	
      \node at (-3.0,12.45)[scale=1] {$t=10\si{\micro\second}$};
      \node at (-3.0, 7.45)[scale=1] {$t=30\si{\micro\second}$};
      \node at (-3.0, 2.45)[scale=1] {$t=60\si{\micro\second}$};
	\end{tikzpicture}
   \caption{N-dodecane shock-droplet interaction calculated with the HLLP (left) and HLLC (right) Riemann solver. The interface is depicted as white line. The top half
   shows a numerical schlieren image, whereas the lower half shows the temperature field.}
   \label{fig:shockdropnull}
\end{figure}

To further highlight the effect of the phase transition process on this test case, we made an additional calculation using the HLLC solver at the interface. A
comparison is shown in Fig.\ref{fig:shockdropnull}, where the left column corresponds to the HLLP solver and the right column to the HLLC solver.  Several
differences to the previously shown results can be seen for the results obtained with the HLLC solver, which does not account for phase transition. For
$t=10\si{\micro\second}$, the bow shock is able to heat up the surrounding gas stronger, due to the absence of an additional shock wave emerging from the
droplet surface. This further leads to slightly faster wave speeds than in the case with phase transition.  For $t=30\si{\micro\second}$, the low temperature region
on the backside of the droplet has a significantly higher temperature than in the HLLP case. This indicates that a large portion of the cooling effect, which is
visible for the HLLP case in this area, can be attributed to the evaporation process. In addition, the instabilities on the droplet surface start to show distinctively
different features. These differences become clear for $t=60\si{\micro\second}$. The deformation  leads to a more elongated droplet perpendicular to the flow
direction. The temperature in the wake shows similar features  than with phase transition, however, less cold vapor is present. Further,the vortex structures and the
detaching shock waves behind the droplet, visible in the schlieren image, show very little similarities with the HLLP case. Concluding, the effect of
phase transition is not negligible for the present shock-drop interaction and influences the deformation of the droplet as well as the wake.

\section{Conclusion}\label{sec:conc}
In this paper we studied  Riemann solvers for the two-phase Riemann problem with phase transition. We obtained closure for the Riemann problem by modelling
phase transition with the theory of classical irreversible thermodynamics. Corner stone of this theory are the Onsager coefficients, which describe the relationship
between the thermodynamic fluxes and their driving forces. We incorporated this modeling strategy in an exact Riemann solver that uses a kinetic relation to
find a unique solution of the two-phase Riemann problem. Further, we formulated two approximate Riemann solvers. They are based on the well-known HLL/HLLC
methodology. One of the Riemann solvers uses a kinetic relation and thus is iterative. The other is purely algebraic and circumvents the need of an iteration
algorithm. The Riemann solvers were employed in a level-set ghost-fluid method, in which the two phase Riemann problem was used to obtain a
thermodynamically consistent coupling between the phases.

We validated our Riemann solvers  against molecular dynamics data of an evaporating Lennard-Jones truncated and shifted fluid. First we used the Riemann solvers
in a modified version of our code, in which the mesh is advected with the phase boundary. Here, a good agreement between sharp-interface method and molecular
dynamics data was achieved for all proposed Riemann solvers. Remaining differences in the vapor temperature near the interface can be attributed to an imperfect
model of the Onsager coefficients. We then tested our full ghost-fluid framework against the reference data and achieved similar results. We  studied the mesh
sensitivity of our solvers and found a tendency for a slight overprediction of the interfacial fluxes for underresolving meshes. Finally, we applied the two
approximate solvers to a complex shock-droplet interaction. The predicted droplet contours were very similar and showed only  slight deviations which were
attributed to a possible underresolution of the droplet surface. Compared with a calculation without phase transition effects, clear differences in the droplet
deformation and flow structures could be observed, highlighting the importance of an accurate description of the interfacial transfer process. 

Future work should follow up on the differences between the HLLP and HLLP0 solver.  It would be ideal to have a purely algebraic interface Riemann solver that
is able to mirror the results of the iterative HLLP solver also in strongly underresolved regions. This might be possible by considering different wave speed
estimates or recovering the contact wave in the approximate Riemann solution. Also, different models for the Onsager coefficients than the one used in this
paper exist in literature \cite{Bedeaux1999,Bond2004,Caputa2011}. It should be evaluated how different thermodynamic models may effect the numerical prediction.
Finally, further applications of the two approximate solvers to complex, multi-dimensional test cases are necessary in order to fully understand the strength
and weaknesses of the iterative and non-iterative approach. 

\section*{Acknowledgements}
This work was supported by the German Research Foundation (DFG) through the Project SFB-TRR 75, Project number 84292822 - ``Droplet Dynamics under Extreme
Ambient Conditions'' and Germany's Excellence Strategy - EXC 2075 – 390740016. The simulations were performed on the national supercomputer HPE Apollo (Hawk) at
the High-Performance Computing Center Stuttgart (HLRS) under the grant numbers \emph{hpcmphas/44084}.

\section*{References}

  \bibliographystyle{elsarticle-num-names} 
  \bibliography{literature}

\begin{thebibliography}{63}
\expandafter\ifx\csname natexlab\endcsname\relax\def\natexlab#1{#1}\fi
\providecommand{\url}[1]{\texttt{#1}}
\providecommand{\href}[2]{#2}
\providecommand{\path}[1]{#1}
\providecommand{\DOIprefix}{doi:}
\providecommand{\ArXivprefix}{arXiv:}
\providecommand{\URLprefix}{URL: }
\providecommand{\Pubmedprefix}{pmid:}
\providecommand{\doi}[1]{\href{http://dx.doi.org/#1}{\path{#1}}}
\providecommand{\Pubmed}[1]{\href{pmid:#1}{\path{#1}}}
\providecommand{\bibinfo}[2]{#2}
\ifx\xfnm\relax \def\xfnm[#1]{\unskip,\space#1}\fi
%Type = Incollection
\bibitem[{J{\"{o}}ns et~al.(2021)J{\"{o}}ns, M{\"{u}}ller, Zeifang, and
  Munz}]{Jons2021}
\bibinfo{author}{S.~J{\"{o}}ns}, \bibinfo{author}{C.~M{\"{u}}ller},
  \bibinfo{author}{J.~Zeifang}, \bibinfo{author}{C.-D. Munz},
\newblock \bibinfo{title}{{Recent Advances and Complex Applications of the
  Compressible Ghost-Fluid Method}},
\newblock \bibinfo{year}{2021}, pp. \bibinfo{pages}{155--176}.
  \DOIprefix\doi{10.1007/978-3-030-72850-2_7}.
%Type = Book
\bibitem[{M{\"{u}}ller et~al.(2020)M{\"{u}}ller, Hitz, J{\"{o}}ns, Zeifang,
  Chiocchetti, and Munz}]{Muller2020}
\bibinfo{author}{C.~M{\"{u}}ller}, \bibinfo{author}{T.~Hitz},
  \bibinfo{author}{S.~J{\"{o}}ns}, \bibinfo{author}{J.~Zeifang},
  \bibinfo{author}{S.~Chiocchetti}, \bibinfo{author}{C.-D. Munz},
  \bibinfo{title}{{Improvement of the Level-Set Ghost-Fluid Method for the
  Compressible Euler Equations}}, volume \bibinfo{volume}{121},
  \bibinfo{year}{2020}. \DOIprefix\doi{10.1007/978-3-030-33338-6_2}.
%Type = Article
\bibitem[{Fechter et~al.(2017)Fechter, Munz, Rohde, and Zeiler}]{Fechter2017}
\bibinfo{author}{S.~Fechter}, \bibinfo{author}{C.~D. Munz},
  \bibinfo{author}{C.~Rohde}, \bibinfo{author}{C.~Zeiler},
\newblock \bibinfo{title}{{A sharp interface method for compressible
  liquid–vapor flow with phase transition and surface tension}},
\newblock \bibinfo{journal}{Journal of Computational Physics}
  \bibinfo{volume}{336} (\bibinfo{year}{2017}) \bibinfo{pages}{347--374}.
%Type = Book
\bibitem[{Ishii and Hibiki(2011)}]{Ishii2011}
\bibinfo{author}{M.~Ishii}, \bibinfo{author}{T.~Hibiki},
  \bibinfo{title}{{Thermo-fluid dynamics of two-phase flow (Second edition)}},
  \bibinfo{year}{2011}. \DOIprefix\doi{10.1007/978-1-4419-7985-8}.
%Type = Article
\bibitem[{Sussman et~al.(1994)Sussman, Smereka, and Osher}]{Sussman1994}
\bibinfo{author}{M.~Sussman}, \bibinfo{author}{P.~Smereka},
  \bibinfo{author}{S.~Osher},
\newblock \bibinfo{title}{{A Level Set Approach for Computing Solutions to
  Incompressible Two-Phase Flow}},
\newblock \bibinfo{journal}{Journal of Computational Physics}
  \bibinfo{volume}{114} (\bibinfo{year}{1994}) \bibinfo{pages}{146--159}.
%Type = Article
\bibitem[{Fedkiw et~al.(1999)Fedkiw, Aslam, Merriman, and Osher}]{Fedkiw1999}
\bibinfo{author}{R.~P. Fedkiw}, \bibinfo{author}{T.~Aslam},
  \bibinfo{author}{B.~Merriman}, \bibinfo{author}{S.~Osher},
\newblock \bibinfo{title}{A non-oscillatory {E}ulerian approach to interfaces
  in multimaterial flows (the ghost fluid method)},
\newblock \bibinfo{journal}{Journal of Computational Physics}
  \bibinfo{volume}{152} (\bibinfo{year}{1999}) \bibinfo{pages}{457--492}.
%Type = Article
\bibitem[{Merkle and Rohde(2007)}]{Merkle2007}
\bibinfo{author}{C.~Merkle}, \bibinfo{author}{C.~Rohde},
\newblock \bibinfo{title}{The sharp-interface approach for fluids with phase
  change: Riemann problems and ghost fluid techniques},
\newblock \bibinfo{journal}{{ESAIM}: Mathematical Modelling and Numerical
  Analysis} \bibinfo{volume}{41} (\bibinfo{year}{2007})
  \bibinfo{pages}{1089--1123}.
%Type = Article
\bibitem[{Fechter et~al.(2018)Fechter, Munz, Rohde, and Zeiler}]{Fechter2018}
\bibinfo{author}{S.~Fechter}, \bibinfo{author}{C.~D. Munz},
  \bibinfo{author}{C.~Rohde}, \bibinfo{author}{C.~Zeiler},
\newblock \bibinfo{title}{{Approximate Riemann solver for compressible liquid
  vapor flow with phase transition and surface tension}},
\newblock \bibinfo{journal}{Computers and Fluids} \bibinfo{volume}{169}
  (\bibinfo{year}{2018}) \bibinfo{pages}{169--185}.
%Type = Article
\bibitem[{Godunov(1959)}]{Godunov1959}
\bibinfo{author}{S.~Godunov},
\newblock \bibinfo{title}{{Finite difference method for numerical computation
  of discontinuous solutions of the equations of fluid dynamics}},
\newblock \bibinfo{journal}{Matematicheskii Sbornik} \bibinfo{volume}{47(89)}
  (\bibinfo{year}{1959}) \bibinfo{pages}{271--306}.
%Type = Book
\bibitem[{Toro(2009)}]{Toro_2009}
\bibinfo{author}{E.~F. Toro}, \bibinfo{title}{{Riemann solvers and numerical
  methods for fluid dynamics: A practical introduction}},
  \bibinfo{publisher}{Springer Berlin Heidelberg}, \bibinfo{year}{2009}.
  \DOIprefix\doi{10.1007/b79761}.
%Type = Article
\bibitem[{Menikoff and Plohr(1989)}]{Menikoff_1989}
\bibinfo{author}{R.~Menikoff}, \bibinfo{author}{B.~J. Plohr},
\newblock \bibinfo{title}{{The Riemann problem for fluid flow of real
  materials}},
\newblock \bibinfo{journal}{Reviews of Modern Physics} \bibinfo{volume}{61}
  (\bibinfo{year}{1989}) \bibinfo{pages}{75--130}.
%Type = Article
\bibitem[{Saurel et~al.(2008)Saurel, Petitpas, and Abgrall}]{Saurel2008}
\bibinfo{author}{R.~Saurel}, \bibinfo{author}{F.~Petitpas},
  \bibinfo{author}{R.~Abgrall},
\newblock \bibinfo{title}{{Modelling phase transition in metastable liquids:
  Application to cavitating and flashing flows}},
\newblock \bibinfo{journal}{Journal of Fluid Mechanics} \bibinfo{volume}{607}
  (\bibinfo{year}{2008}) \bibinfo{pages}{313--350}.
%Type = Article
\bibitem[{{Le M{\'{e}}tayer} et~al.(2013){Le M{\'{e}}tayer}, Massoni, and
  Saurel}]{LeMetayer2013}
\bibinfo{author}{O.~{Le M{\'{e}}tayer}}, \bibinfo{author}{J.~Massoni},
  \bibinfo{author}{R.~Saurel},
\newblock \bibinfo{title}{{Dynamic relaxation processes in compressible
  multiphase flows. Application to evaporation phenomena}},
\newblock \bibinfo{journal}{ESAIM: Proceedings} \bibinfo{volume}{40}
  (\bibinfo{year}{2013}) \bibinfo{pages}{103--123}.
%Type = Article
\bibitem[{Furfaro and Saurel(2015)}]{Furfaro2015}
\bibinfo{author}{D.~Furfaro}, \bibinfo{author}{R.~Saurel},
\newblock \bibinfo{title}{{A simple HLLC-type Riemann solver for compressible
  non-equilibrium two-phase flows}},
\newblock \bibinfo{journal}{Computers and Fluids} \bibinfo{volume}{111}
  (\bibinfo{year}{2015}) \bibinfo{pages}{159--178}.
%Type = Article
\bibitem[{Kuila et~al.(2015)Kuila, {Raja Sekhar}, and Zeidan}]{Kuila2015}
\bibinfo{author}{S.~Kuila}, \bibinfo{author}{T.~{Raja Sekhar}},
  \bibinfo{author}{D.~Zeidan},
\newblock \bibinfo{title}{{A Robust and accurate Riemann solver for a
  compressible two-phase flow model}},
\newblock \bibinfo{journal}{Applied Mathematics and Computation}
  \bibinfo{volume}{265} (\bibinfo{year}{2015}) \bibinfo{pages}{681--695}.
%Type = Article
\bibitem[{Schwendeman et~al.(2006)Schwendeman, Wahle, and
  Kapila}]{Schwendeman2006}
\bibinfo{author}{D.~W. Schwendeman}, \bibinfo{author}{C.~W. Wahle},
  \bibinfo{author}{A.~K. Kapila},
\newblock \bibinfo{title}{{The Riemann problem and a high-resolution Godunov
  method for a model of compressible two-phase flow}},
\newblock \bibinfo{journal}{Journal of Computational Physics}
  \bibinfo{volume}{212} (\bibinfo{year}{2006}) \bibinfo{pages}{490--526}.
%Type = Article
\bibitem[{Abeyaratne and Knowles(1991)}]{Abeyaratne1991}
\bibinfo{author}{R.~Abeyaratne}, \bibinfo{author}{J.~K. Knowles},
\newblock \bibinfo{title}{{Kinetic relations and the propagation of phase
  boundaries in solids}},
\newblock \bibinfo{journal}{Archive for Rational Mechanics and Analysis}
  \bibinfo{volume}{114} (\bibinfo{year}{1991}) \bibinfo{pages}{119--154}.
%Type = Article
\bibitem[{{Le M{\'{e}}tayer} et~al.(2005){Le M{\'{e}}tayer}, Massoni, and
  Saurel}]{LeMetayer2005}
\bibinfo{author}{O.~{Le M{\'{e}}tayer}}, \bibinfo{author}{J.~Massoni},
  \bibinfo{author}{R.~Saurel},
\newblock \bibinfo{title}{{Modelling evaporation fronts with reactive Riemann
  solvers}},
\newblock \bibinfo{journal}{Journal of Computational Physics}
  \bibinfo{volume}{205} (\bibinfo{year}{2005}) \bibinfo{pages}{567--610}.
%Type = Article
\bibitem[{Hantke and Thein(2019)}]{Hantke2019}
\bibinfo{author}{M.~Hantke}, \bibinfo{author}{F.~Thein},
\newblock \bibinfo{title}{{On the Impossibility of First-Order Phase
  Transitions in Systems Modeled by the Full Euler Equations}},
\newblock \bibinfo{journal}{Entropy} \bibinfo{volume}{21}
  (\bibinfo{year}{2019}) \bibinfo{pages}{1039}.
%Type = Article
\bibitem[{Rohde and Zeiler(2015)}]{Rohde2015}
\bibinfo{author}{C.~Rohde}, \bibinfo{author}{C.~Zeiler},
\newblock \bibinfo{title}{{A relaxation Riemann solver for compressible
  two-phase flow with phase transition and surface tension}},
\newblock \bibinfo{journal}{Applied Numerical Mathematics} \bibinfo{volume}{95}
  (\bibinfo{year}{2015}) \bibinfo{pages}{267--279}.
%Type = Article
\bibitem[{Hantke et~al.(2013)Hantke, Dreyer, and Warnecke}]{Hantke_2013}
\bibinfo{author}{M.~Hantke}, \bibinfo{author}{W.~Dreyer},
  \bibinfo{author}{G.~Warnecke},
\newblock \bibinfo{title}{{Exact solutions to the Riemann problem for
  compressible isothermal Euler equations for two-phase flows with and without
  phase transition}},
\newblock \bibinfo{journal}{Quarterly of Applied Mathematics}
  \bibinfo{volume}{71} (\bibinfo{year}{2013}) \bibinfo{pages}{509--540}.
%Type = Phdthesis
\bibitem[{Thein(2018)}]{Thein2018}
\bibinfo{author}{F.~Thein}, \bibinfo{title}{{Results for Two Phase Flows with
  Phase Transition}}, Ph.D. thesis, Otto-von-Guericke-Universität Magdeburg,
  \bibinfo{year}{2018}.
%Type = Article
\bibitem[{Hitz et~al.(2021)Hitz, J{\"{o}}ns, Heinen, Vrabec, and
  Munz}]{Hitz2020}
\bibinfo{author}{T.~Hitz}, \bibinfo{author}{S.~J{\"{o}}ns},
  \bibinfo{author}{M.~Heinen}, \bibinfo{author}{J.~Vrabec},
  \bibinfo{author}{C.-D. Munz},
\newblock \bibinfo{title}{{Comparison of macro- and microscopic solutions of
  the Riemann problem II. Two-phase shock tube}},
\newblock \bibinfo{journal}{Journal of Computational Physics}
  \bibinfo{volume}{429} (\bibinfo{year}{2021}) \bibinfo{pages}{110027}.
%Type = Phdthesis
\bibitem[{M\"uller(2021)}]{Mueller2021}
\bibinfo{author}{C.~M\"uller}, \bibinfo{title}{{Multiscale Modeling of the
  Evaporation Process}}, Ph.D. thesis, University of Stuttgart,
  \bibinfo{year}{2021}.
%Type = Article
\bibitem[{F{\"{o}}ll et~al.(2019)F{\"{o}}ll, Hitz, M{\"{u}}ller, Munz, and
  Dumbser}]{Foll2019}
\bibinfo{author}{F.~F{\"{o}}ll}, \bibinfo{author}{T.~Hitz},
  \bibinfo{author}{C.~M{\"{u}}ller}, \bibinfo{author}{C.~D. Munz},
  \bibinfo{author}{M.~Dumbser},
\newblock \bibinfo{title}{{On the use of tabulated equations of state for
  multi-phase simulations in the homogeneous equilibrium limit}},
\newblock \bibinfo{journal}{Shock Waves} \bibinfo{volume}{29}
  (\bibinfo{year}{2019}) \bibinfo{pages}{769--793}.
%Type = Book
\bibitem[{Gyarmati(1970)}]{Gyarmati1970}
\bibinfo{author}{I.~Gyarmati}, \bibinfo{title}{{Non-equilibrium
  Thermodynamics}}, Ingenieurwissenschaftliche Bibliothek / Engineering Science
  Library, \bibinfo{publisher}{Springer Berlin Heidelberg},
  \bibinfo{address}{Berlin, Heidelberg}, \bibinfo{year}{1970}.
  \DOIprefix\doi{10.1007/978-3-642-51067-0}.
%Type = Book
\bibitem[{de~Groot and Mazur(1984)}]{deGroot1984}
\bibinfo{author}{S.~R. de~Groot}, \bibinfo{author}{P.~Mazur},
  \bibinfo{title}{Non-equilibrium Thermodynamics}, \bibinfo{publisher}{Dover
  Publications}, \bibinfo{address}{New York}, \bibinfo{year}{1984}.
%Type = Book
\bibitem[{Lebon et~al.(2008)Lebon, Jou, and Casas-V{\'{a}}zquez}]{Lebon2008}
\bibinfo{author}{G.~Lebon}, \bibinfo{author}{D.~Jou},
  \bibinfo{author}{J.~Casas-V{\'{a}}zquez}, \bibinfo{title}{{Understanding
  Non-equilibrium Thermodynamics}}, \bibinfo{publisher}{Springer Berlin
  Heidelberg}, \bibinfo{address}{Berlin, Heidelberg}, \bibinfo{year}{2008}.
  \DOIprefix\doi{10.1007/978-3-540-74252-4}.
%Type = Book
\bibitem[{{S. Kjelstrup and D. Bedeaux}(2008)}]{Kjelstrup2008}
\bibinfo{author}{{S. Kjelstrup and D. Bedeaux}},
  \bibinfo{title}{{Non-equilibrium thermodynamics of heterogeneous systems}},
  \bibinfo{edition}{16} ed., \bibinfo{publisher}{WORLD SCIENTIFIC},
  \bibinfo{year}{2008}.
%Type = Article
\bibitem[{Cipolla et~al.(1974)Cipolla, Lang, and Loyalka}]{Cipolla1974}
\bibinfo{author}{J.~W. Cipolla}, \bibinfo{author}{H.~Lang},
  \bibinfo{author}{S.~K. Loyalka},
\newblock \bibinfo{title}{{Kinetic theory of condensation and evaporation.
  II}},
\newblock \bibinfo{journal}{The Journal of Chemical Physics}
  \bibinfo{volume}{61} (\bibinfo{year}{1974}) \bibinfo{pages}{69--77}.
%Type = Article
\bibitem[{Johannessen and Bedeaux(2006)}]{Johannessen2006}
\bibinfo{author}{E.~Johannessen}, \bibinfo{author}{D.~Bedeaux},
\newblock \bibinfo{title}{{Integral relations for the heat and mass transfer
  resistivities of the liquid-vapor interface}},
\newblock \bibinfo{journal}{Physica A: Statistical Mechanics and its
  Applications} \bibinfo{volume}{370} (\bibinfo{year}{2006})
  \bibinfo{pages}{258--274}.
%Type = Article
\bibitem[{Stierle et~al.(2020)Stierle, Waibel, Gross, Steinhausen, Weigand, and
  Lamanna}]{Stierle2020}
\bibinfo{author}{R.~Stierle}, \bibinfo{author}{C.~Waibel},
  \bibinfo{author}{J.~Gross}, \bibinfo{author}{C.~Steinhausen},
  \bibinfo{author}{B.~Weigand}, \bibinfo{author}{G.~Lamanna},
\newblock \bibinfo{title}{{On the Selection of Boundary Conditions for Droplet
  Evaporation and Condensation at high Pressure and Temperature Conditions from
  interfacial Transport Resistivities}},
\newblock \bibinfo{journal}{International Journal of Heat and Mass Transfer}
  \bibinfo{volume}{151} (\bibinfo{year}{2020}) \bibinfo{pages}{119450}.
%Type = Article
\bibitem[{Nagayama et~al.(2015)Nagayama, Takematsu, Mizuguchi, and
  Tsuruta}]{Nagayama2015}
\bibinfo{author}{G.~Nagayama}, \bibinfo{author}{M.~Takematsu},
  \bibinfo{author}{H.~Mizuguchi}, \bibinfo{author}{T.~Tsuruta},
\newblock \bibinfo{title}{{Molecular dynamics study on condensation/evaporation
  coefficients of chain molecules at liquid-vapor interface}},
\newblock \bibinfo{journal}{Journal of Chemical Physics} \bibinfo{volume}{143}
  (\bibinfo{year}{2015}).
%Type = Phdthesis
\bibitem[{Fechter(2015)}]{Fechter2015b}
\bibinfo{author}{S.~D. Fechter}, \bibinfo{title}{{Compressible multi-phase
  simulation at extreme conditions using a discontinuous Galerkin scheme}},
  Ph.D. thesis, University of Stuttgart, \bibinfo{year}{2015}.
%Type = Phdthesis
\bibitem[{Zeifang(2020)}]{Zeifang2020}
\bibinfo{author}{J.~Zeifang}, \bibinfo{title}{{A Discontinuous Galerkin Method
  for Droplet Dynamics in Weakly Compressible Flows}}, Ph.D. thesis, University
  of Stuttgart, \bibinfo{year}{2020}.
%Type = Article
\bibitem[{Krais et~al.(2020)Krais, Beck, Bolemann, Frank, Flad, Gassner,
  Hindenlang, Hoffmann, Kuhn, Sonntag, and Munz}]{Krais2020}
\bibinfo{author}{N.~Krais}, \bibinfo{author}{A.~Beck},
  \bibinfo{author}{T.~Bolemann}, \bibinfo{author}{H.~Frank},
  \bibinfo{author}{D.~Flad}, \bibinfo{author}{G.~Gassner},
  \bibinfo{author}{F.~Hindenlang}, \bibinfo{author}{M.~Hoffmann},
  \bibinfo{author}{T.~Kuhn}, \bibinfo{author}{M.~Sonntag},
  \bibinfo{author}{C.-D. Munz},
\newblock \bibinfo{title}{{FLEXI: A high order discontinuous Galerkin framework
  for hyperbolic–parabolic conservation laws}},
\newblock \bibinfo{journal}{Computers {\&} Mathematics with Applications}
  (\bibinfo{year}{2020}).
%Type = Article
\bibitem[{Bassi et~al.(2011)Bassi, Franchina, Ghidoni, and Rebay}]{Bassi2011}
\bibinfo{author}{F.~Bassi}, \bibinfo{author}{N.~Franchina},
  \bibinfo{author}{A.~Ghidoni}, \bibinfo{author}{S.~Rebay},
\newblock \bibinfo{title}{{Spectral p-multigrid discontinuous Galerkin solution
  of the Navier-Stokes equations}},
\newblock \bibinfo{journal}{International Journal for Numerical Methods in
  Fluids} \bibinfo{volume}{67} (\bibinfo{year}{2011})
  \bibinfo{pages}{1540--1558}.
%Type = Article
\bibitem[{Bassi and Rebay(2002)}]{Bassi2002}
\bibinfo{author}{F.~Bassi}, \bibinfo{author}{S.~Rebay},
\newblock \bibinfo{title}{{Numerical evaluation of two discontinuous Galerkin
  methods for the compressible Navier-Stokes equations}},
\newblock \bibinfo{journal}{International Journal for Numerical Methods in
  Fluids} \bibinfo{volume}{40} (\bibinfo{year}{2002})
  \bibinfo{pages}{197--207}.
%Type = Inproceedings
\bibitem[{Sonntag and Munz(2014)}]{Sonntag2014}
\bibinfo{author}{M.~Sonntag}, \bibinfo{author}{C.~D. Munz},
\newblock \bibinfo{title}{{Shock capturing for discontinuous galerkin methods
  using finite volume subcells}},
\newblock in: \bibinfo{booktitle}{Springer Proceedings in Mathematics and
  Statistics}, \bibinfo{year}{2014}.
  \DOIprefix\doi{10.1007/978-3-319-05591-6_96}.
%Type = Article
\bibitem[{Sonntag and Munz(2017)}]{Sonntag2017}
\bibinfo{author}{M.~Sonntag}, \bibinfo{author}{C.~D. Munz},
\newblock \bibinfo{title}{{Efficient Parallelization of a Shock Capturing for
  Discontinuous Galerkin Methods using Finite Volume Sub-cells}},
\newblock \bibinfo{journal}{Journal of Scientific Computing}
  \bibinfo{volume}{70} (\bibinfo{year}{2017}) \bibinfo{pages}{1262--1289}.
%Type = Inproceedings
\bibitem[{Persson and Peraire(2006)}]{Persson2006}
\bibinfo{author}{P.-O. Persson}, \bibinfo{author}{J.~Peraire},
\newblock \bibinfo{title}{{Sub-Cell Shock Capturing for Discontinuous Galerkin
  Methods}},
\newblock in: \bibinfo{booktitle}{44th AIAA Aerospace Sciences Meeting and
  Exhibit}, \bibinfo{publisher}{American Institute of Aeronautics and
  Astronautics}, \bibinfo{address}{Reston, Virigina}, \bibinfo{year}{2006}.
  \DOIprefix\doi{10.2514/6.2006-112}.
%Type = Techreport
\bibitem[{Carpenter and Kennedy(1994)}]{Carpenter1994}
\bibinfo{author}{M.~Carpenter}, \bibinfo{author}{C.~Kennedy},
  \bibinfo{title}{Fourth-order {$2N$}-storage {Runge-Kutta} schemes},
  \bibinfo{type}{Technical Report}, NASA Langley Research Center,
  \bibinfo{year}{1994}.
%Type = Article
\bibitem[{Peng et~al.(1999)Peng, Merriman, Osher, Zhao, and Kang}]{Peng1999}
\bibinfo{author}{D.~Peng}, \bibinfo{author}{B.~Merriman},
  \bibinfo{author}{S.~Osher}, \bibinfo{author}{H.~Zhao},
  \bibinfo{author}{M.~Kang},
\newblock \bibinfo{title}{{A PDE-Based Fast Local Level Set Method}},
\newblock \bibinfo{journal}{Journal of Computational Physics}
  \bibinfo{volume}{155} (\bibinfo{year}{1999}) \bibinfo{pages}{410--438}.
%Type = Article
\bibitem[{Jiang and Peng(2000)}]{Jiang2000}
\bibinfo{author}{G.~S. Jiang}, \bibinfo{author}{D.~Peng},
\newblock \bibinfo{title}{{Weighted ENO schemes for Hamilton-Jacobi
  equations}},
\newblock \bibinfo{journal}{SIAM Journal on Scientific Computing}
  \bibinfo{volume}{21} (\bibinfo{year}{2000}) \bibinfo{pages}{2126--2143}.
%Type = Article
\bibitem[{Castro et~al.(2006)Castro, Gallardo, and Par{\'e}s}]{castro2006high}
\bibinfo{author}{M.~Castro}, \bibinfo{author}{J.~Gallardo},
  \bibinfo{author}{C.~Par{\'e}s},
\newblock \bibinfo{title}{High order finite volume schemes based on
  reconstruction of states for solving hyperbolic systems with nonconservative
  products. applications to shallow-water systems},
\newblock \bibinfo{journal}{Mathematics of computation} \bibinfo{volume}{75}
  (\bibinfo{year}{2006}) \bibinfo{pages}{1103--1134}.
%Type = Article
\bibitem[{Dumbser and Loub{\`{e}}re(2016)}]{Dumbser2016a}
\bibinfo{author}{M.~Dumbser}, \bibinfo{author}{R.~Loub{\`{e}}re},
\newblock \bibinfo{title}{A simple robust and accurate a posteriori sub-cell
  finite volume limiter for the discontinuous {G}alerkin method on unstructured
  meshes},
\newblock \bibinfo{journal}{Journal of Computational Physics}
  \bibinfo{volume}{319} (\bibinfo{year}{2016}) \bibinfo{pages}{163--199}.
%Type = Phdthesis
\bibitem[{Hitz(2020)}]{Hitz2020a}
\bibinfo{author}{T.~Hitz}, \bibinfo{title}{{On the Riemann Problem and the
  Navier-Stokes-Korteweg Model for Compressible Multiphase Flows}}, Ph.D.
  thesis, University of Stuttgart, \bibinfo{year}{2020}.
%Type = Article
\bibitem[{Minoli and Kopriva(2011)}]{Minoli2011}
\bibinfo{author}{C.~A. Minoli}, \bibinfo{author}{D.~A. Kopriva},
\newblock \bibinfo{title}{{Discontinuous Galerkin spectral element
  approximations on moving meshes}},
\newblock \bibinfo{journal}{Journal of Computational Physics}
  \bibinfo{volume}{230} (\bibinfo{year}{2011}) \bibinfo{pages}{1876--1902}.
%Type = Phdthesis
\bibitem[{Zeiler(2015)}]{Zeiler2015}
\bibinfo{author}{C.~Zeiler}, \bibinfo{title}{{Liquid Vapor Phase Transitions:
  Modeling, Riemann Solvers and Computation}}, Ph.D. thesis, University of
  Stuttgart, \bibinfo{year}{2015}.
%Type = Manual
\bibitem[{Galassi et~al.(2009)Galassi, Davies, Theiler, Gough, Jungman, Alken,
  Booth, Fabrice, and Ulerich}]{Galassi}
\bibinfo{author}{M.~Galassi}, \bibinfo{author}{J.~Davies},
  \bibinfo{author}{J.~Theiler}, \bibinfo{author}{B.~Gough},
  \bibinfo{author}{G.~Jungman}, \bibinfo{author}{P.~Alken},
  \bibinfo{author}{M.~Booth}, \bibinfo{author}{R.~Fabrice},
  \bibinfo{author}{R.~Ulerich}, \bibinfo{title}{{GNU Scientific Library
  Reference Manual }}, \bibinfo{year}{2009}.
%Type = Article
\bibitem[{Harten et~al.(1983)Harten, Lax, and van Leer}]{Harten_1983}
\bibinfo{author}{A.~Harten}, \bibinfo{author}{P.~D. Lax},
  \bibinfo{author}{B.~van Leer},
\newblock \bibinfo{title}{{On Upstream Differencing and Godunov-Type Schemes
  for Hyperbolic Conservation Laws}},
\newblock \bibinfo{journal}{SIAM Review} \bibinfo{volume}{25}
  (\bibinfo{year}{1983}) \bibinfo{pages}{35--61}.
%Type = Article
\bibitem[{Toro et~al.(1994)Toro, Spruce, and Speares}]{Toro_1994}
\bibinfo{author}{E.~F. Toro}, \bibinfo{author}{M.~Spruce},
  \bibinfo{author}{W.~Speares},
\newblock \bibinfo{title}{{Restoration of the contact surface in the
  HLL-Riemann solver}},
\newblock \bibinfo{journal}{Shock Waves} \bibinfo{volume}{4}
  (\bibinfo{year}{1994}) \bibinfo{pages}{25--34}.
%Type = Article
\bibitem[{Hu et~al.(2009)Hu, Adams, and Iaccarino}]{Hu2009}
\bibinfo{author}{X.~Hu}, \bibinfo{author}{N.~Adams},
  \bibinfo{author}{G.~Iaccarino},
\newblock \bibinfo{title}{{On the HLLC Riemann solver for interface interaction
  in compressible multi-fluid flow}},
\newblock \bibinfo{journal}{Journal of Computational Physics}
  \bibinfo{volume}{228} (\bibinfo{year}{2009}) \bibinfo{pages}{6572--6589}.
%Type = Article
\bibitem[{Davis(1988)}]{Davis1988}
\bibinfo{author}{S.~F. Davis},
\newblock \bibinfo{title}{{Simplified Second-Order Godunov-Type Methods}},
\newblock \bibinfo{journal}{SIAM Journal on Scientific and Statistical
  Computing} \bibinfo{volume}{9} (\bibinfo{year}{1988})
  \bibinfo{pages}{445--473}.
%Type = Article
\bibitem[{Heier et~al.(2018)Heier, Stephan, Liu, Chapman, Hasse, and
  Langenbach}]{Heier2018}
\bibinfo{author}{M.~Heier}, \bibinfo{author}{S.~Stephan},
  \bibinfo{author}{J.~Liu}, \bibinfo{author}{W.~G. Chapman},
  \bibinfo{author}{H.~Hasse}, \bibinfo{author}{K.~Langenbach},
\newblock \bibinfo{title}{{Equation of state for the Lennard-Jones truncated
  and shifted fluid with a cut-off radius of 2.5 $\sigma$ based on perturbation
  theory and its applications to interfacial thermodynamics}},
\newblock \bibinfo{journal}{Molecular Physics} \bibinfo{volume}{116}
  (\bibinfo{year}{2018}) \bibinfo{pages}{2083--2094}.
%Type = Article
\bibitem[{Homes et~al.(2021)Homes, Heinen, Vrabec, and Fischer}]{Homes2021}
\bibinfo{author}{S.~Homes}, \bibinfo{author}{M.~Heinen},
  \bibinfo{author}{J.~Vrabec}, \bibinfo{author}{J.~Fischer},
\newblock \bibinfo{title}{{Evaporation driven by conductive heat transport}},
\newblock \bibinfo{journal}{Molecular Physics} \bibinfo{volume}{119}
  (\bibinfo{year}{2021}).
%Type = Article
\bibitem[{Lautenschlaeger and Hasse(2019)}]{Lautenschlaeger2019}
\bibinfo{author}{M.~P. Lautenschlaeger}, \bibinfo{author}{H.~Hasse},
\newblock \bibinfo{title}{{Transport properties of the Lennard-Jones truncated
  and shifted fluid from non-equilibrium molecular dynamics simulations}},
\newblock \bibinfo{journal}{Fluid Phase Equilibria} \bibinfo{volume}{482}
  (\bibinfo{year}{2019}) \bibinfo{pages}{38--47}.
%Type = Article
\bibitem[{Lemmon and Jacobsen(2004)}]{Lemmon2004}
\bibinfo{author}{E.~W. Lemmon}, \bibinfo{author}{R.~T. Jacobsen},
\newblock \bibinfo{title}{{Viscosity and Thermal Conductivity Equations for
  Nitrogen, Oxygen, Argon, and Air}},
\newblock \bibinfo{journal}{International Journal of Thermophysics}
  \bibinfo{volume}{25} (\bibinfo{year}{2004}) \bibinfo{pages}{21--69}.
%Type = Article
\bibitem[{Merker et~al.(2012)Merker, Vrabec, and Hasse}]{Merker2012}
\bibinfo{author}{T.~Merker}, \bibinfo{author}{J.~Vrabec},
  \bibinfo{author}{H.~Hasse},
\newblock \bibinfo{title}{{Engineering molecular models: Efficient
  parameterization procedure and cyclohexanol as case study}},
\newblock \bibinfo{journal}{Soft Materials} \bibinfo{volume}{10}
  (\bibinfo{year}{2012}) \bibinfo{pages}{3--25}.
%Type = Article
\bibitem[{Chung et~al.(1988)Chung, Ajlan, Lee, and Starling}]{Chung1988}
\bibinfo{author}{T.-H. Chung}, \bibinfo{author}{M.~Ajlan},
  \bibinfo{author}{L.~L. Lee}, \bibinfo{author}{K.~E. Starling},
\newblock \bibinfo{title}{{Generalized Multiparameter Correlation for Nonpolar
  and Polar Fluid Transport Properties}},
\newblock \bibinfo{journal}{Ind. Eng. Chem Res.} \bibinfo{volume}{27}
  (\bibinfo{year}{1988}) \bibinfo{pages}{671--679}.
%Type = Article
\bibitem[{Bedeaux and Kjelstrup(1999)}]{Bedeaux1999}
\bibinfo{author}{D.~Bedeaux}, \bibinfo{author}{S.~Kjelstrup},
\newblock \bibinfo{title}{{Transfer coefficients for evaporation}},
\newblock \bibinfo{journal}{Physica A: Statistical Mechanics and its
  Applications} \bibinfo{volume}{270} (\bibinfo{year}{1999})
  \bibinfo{pages}{413--426}.
%Type = Article
\bibitem[{Bond and Struchtrup(2004)}]{Bond2004}
\bibinfo{author}{M.~Bond}, \bibinfo{author}{H.~Struchtrup},
\newblock \bibinfo{title}{{Mean evaporation and condensation coefficients based
  on energy dependent condensation probability}},
\newblock \bibinfo{journal}{Physical Review E - Statistical Physics, Plasmas,
  Fluids, and Related Interdisciplinary Topics} \bibinfo{volume}{70}
  (\bibinfo{year}{2004}) \bibinfo{pages}{21}.
%Type = Article
\bibitem[{Caputa and Struchtrup(2011)}]{Caputa2011}
\bibinfo{author}{J.~P. Caputa}, \bibinfo{author}{H.~Struchtrup},
\newblock \bibinfo{title}{{Interface model for non-equilibrium evaporation}},
\newblock \bibinfo{journal}{Physica A: Statistical Mechanics and its
  Applications} \bibinfo{volume}{390} (\bibinfo{year}{2011})
  \bibinfo{pages}{31--42}.

\end{thebibliography}

\end{document}